\documentclass[twocolumn,english,aps,prb,showpacs,superscriptaddress]{revtex4-1}
\usepackage[T1]{fontenc}
\usepackage[latin9]{inputenc}
\usepackage{textcomp}
\usepackage{amsmath}
\usepackage{amssymb}
\usepackage{graphicx}
\usepackage{color}

\makeatletter

\usepackage{ulem}
\usepackage{float}
\usepackage{lipsum}
\usepackage{babel}

\makeatother

\usepackage{babel}
\begin{document}
\title{Engineering an imaginary stark ladder in a dissipative lattice:  passive $\mathcal{PT}$ symmetry, $K$ symmetry and localized damping}
\author{Yu Zhang}
\affiliation{Beijing National Laboratory for Condensed Matter Physics, Institute
of Physics, Chinese Academy of Sciences, Beijing 100190, China}
\affiliation{School of Physical Sciences, University of Chinese Academy of Sciences,
Beijing 100049, China }
\author{Shu Chen}
\email{Corresponding author: schen@iphy.ac.cn }
\affiliation{Beijing National Laboratory for Condensed Matter Physics, Institute
of Physics, Chinese Academy of Sciences, Beijing 100190, China}
\affiliation{School of Physical Sciences, University of Chinese Academy of Sciences,
Beijing 100049, China }
\date{\today}
\begin{abstract}
We study an imaginary stark ladder model
and propose a realization of the model in a dissipative chain with linearly increasing site-dependent dissipation strength.  Due to the existence of a $K$-symmetry and passive $\mathcal{PT}$ symmetry, the model exhibits quite different feature from its Hermitian counterpart.
With the increase of dissipation strength, the system first undergoes a passive $\mathcal{PT}$-symmetry breaking transition, with the shifted eigenvalues changing from real to complex, and then a $K$-symmetry restoring transition, characterized by the emergence of pure imaginary spectrum with equal spacing.
Accordingly, the eigenstates change from $\mathcal{PT}$-unbroken extended states to the $\mathcal{PT}$-broken states, and finally to stark localized states.
In the framework of the quantum open system governed by Lindblad equation with linearly increasing site-dependent dissipation, we unveil that the dynamical evolution of single particle correlation function is governed by the Hamiltonian of the imaginary stark ladder model. By studying the dynamical evolution of the density distribution under various initial states, we demonstrate that the damping dynamics displays distinct behaviors in different regions. A localized damping is observed in the strong dissipation limit.
\end{abstract}
\maketitle

\section{Introduction}
Advances in manipulating dissipation in laboratory have led to a renewed
interest in the study of open quantum systems. Dissipative processes have been
employed as a tool to engineer quantum states \cite{Muller,Cirac2009,Barreiro}, e.g., a dissipative
coupling was used to experimentally realize a Tonks-Girardeau gas of
molecules \cite{Syassen}, achieve topological states \cite{Diehl2011}, and study peculiar dynamical behaviors associated with passive $\mathcal{PT}$ symmetry in dissipative quantum systems \cite{LuoL,ZhangW,XueP,Naghiloo}. The feasibility of simulating non-Hermitian Hamiltonians in
the context of open quantum systems attracts a growing interest in studying unique features with no counterpart in Hermitian systems, for example, the non-Hermitian skin effect \cite{TELee,SYao1,Kunst,LeeCH,Alvarez}, enriched symmetries \cite{BL} and topological classifications beyond the standard ten classes \cite{Gong,HYZhou,Sato,LiuCH-2019,LiuCH-PRB}. Some recent works have unveiled the unique dynamical signatures of the non-Hermitian
skin effect in dissipative systems \cite{skin effect dissipative,information constraint}, the emergence of chiral damping \cite{chiraldamping} and helical damping \cite{helicaldamping}, and edge burst \cite{WangZ2022}.
Experimental observation of dynamical signature of  non-Hermitian skin effect has been recently reported in the ultracold atomic gases \cite{YanB}.

Quantum simulation in virtue of controllable dissipations greatly expands the research field of traditional condensed matter systems, which stimulates us to explore novel effect induced by purely imaginary potentials \cite{YanB,LiuYG,YiW}. In this work, we shall focus on the problem of particles on a lattice with a linearly increasing imaginary potential. It is well known that an electron in a crystal subjected to an electric field can be described by a lattice model with linear potential, known as the stark ladder model, which yields an equispaced energy spectrum and localized eigenstates controlled by the electric field strength \citep{Wannier,stark_localization_wainner}. The stark ladder model can be also realized in tilted optical lattice \cite{Dahan,Raizen}. This model has been extensively studied \citep{stark_localization_review,stark-localization-solve} with particular attention on interesting phenomena, like Bloch oscillations \citep{Bloch oscillation}, Zener tunneling \citep{Zener tunneling}, dynamical localization driven by oscillating fields \citep{stark ac+dc}, and many-body stark localization \citep{stark-MBL-theory,stark-MBL-experiment}.

When the dissipation effect is considered, early works mainly focused
on the fate of the Bloch oscillation \citep{Bloch-oscillation PT symmetry}, Zener
tunneling \citep{Chiral Zener} and dynamical localization in complex
crystals \citep{dynamic localization complex crystals}. A recent work
studied the interplay between non-Hermitian skin effect and electric
fields \citep{non-hermitian skin effect+stark}. However, the effect of an imaginary linear potential induced by dissipation is not explored yet. It is the aim of this work
to investigate the imaginary stark ladder model in a dissipation lattice, highlighting some unique features arising from the purely imaginary linear potential.

In contrast to the stark ladder model which has no symmetry, an imaginary stark ladder model possesses a $K$-symmetry and passive $\mathcal{PT}$ symmetry, which render
the spectrum displaying quite different features from its Hermitian counterpart.  When the strength of imaginary potential exceeds a threshold, a passive $\mathcal{PT}$ symmetry breaking occurs and the shifted eigenvalues become complex. When we further increase the strength to exceed another threshold, the $K$ symmetry is restored as all eigenvalues lie in the imaginary axis.
We then explore the physical realization of the imaginary stark ladder model via the engineering
of dissipative lattices, for which key characters of
imaginary stark ladder are  built into the
dissipative quantum jump processes of a Lindblad master
equation \citep{chiraldamping,information constraint,skin effect dissipative,helicaldamping}.

Our paper is organized as follows. We introduce our model in Sec.\ref{part:Model}
and analyze the symmetry of the system.
Then we study the $\mathcal{PT}$ symmetry breaking and $K$ symmetry restoration when we increase the strength of imaginary potential.
We also discuss the effects of finite system size and open boundary condition.
In Sec.\ref{part:open system},
we propose how to realize this model in the open quantum system with linearly increasing on-site dissipation and  demonstrate the unique dynamical properties of the model.
A summary is given in Sec.\ref{part:Conclusion-and-discussion}.

\section{Model, symmetries and spectrum \label{part:Model}}

We consider a one-dimensional lattice model with linear imaginary potentials described by the Hamiltonian:
\begin{equation}
\hat{H}=\frac{J}{2}\sum_{j=1}^{L-1}(\hat{c}^{\dagger}_{j+1}\hat{c}_j+h.c.)-iF\sum_{j=1}^{L}j\hat{n}_j,  \label{eq:Hamiltonian}
\end{equation}
where $\hat{c}_j(\hat{c}_j^{\dagger})$ is the fermionic annihilation (creation) operator on the $j_{th}$ site, $\hat{n}_j$ is the particle number operator on the $j_{th}$ site,
 $J/2$ is the strength of the hopping term between neighboring sites, $F$ denotes the strength of the linear imaginary potential,
and $L$ is the number of sites on the lattice with the boundary conditions taken to be open.
The model (\ref{eq:Hamiltonian}) is distinguished  from the stark (or Wannier-stark) ladder model \citep{Wannier} by the potential being imaginary. For convenience, we coin the  model (\ref{eq:Hamiltonian}) as the imaginary stark ladder model. In the following calculation, we shall take $J=1$ as the unit of energy.

Although the imaginary stark ladder model has a very similar form to the stark ladder model, it displays quite different spectrum structure from the stark ladder model due to the existence of two symmetries. At first, we indicate that the Hamiltonian (\ref{eq:Hamiltonian}) after making an overall energy shift fulfills a $K$ symmetry \cite{BL}.
Explicitly, the Hamiltonian with an overall energy shift is written as
\begin{equation}
\hat{H'}=\hat{H}+ i\frac{L+1}{2}F(\sum_j \hat{c}_{j}^{\dagger} \hat{c}_{j}),
\end{equation}
which fulfills:
\begin{equation}
\hat{U}^{-1}\hat{H'}\hat{U}=\hat{H'}^{\dagger}. \label{K-symmetry1}
\end{equation}
Here $\hat{U}$ corresponds to a transformation:
\[
\hat{c}_{j}\rightarrow (-1)^{j}\hat{c}_{j}^\dagger.
\]
Note that $\hat{H'}$ can be represented as
$$
\hat{H'}=\textbf{c}^\dagger H' \textbf{c},
$$
where $\textbf{c}=(\hat{c}_1, \hat{c}_2\dots \hat{c}_L)^T$ , $\textbf{c}^\dagger=(\hat{c}_1^\dagger, \hat{c}_2^\dagger\dots \hat{c}_L^\dagger)$ and $H'$ is the matrix form of the Hamiltonian $\hat{H'}$. Correspondingly, we use $H$ to represent the matrix form of $\hat{H}$. Then Eq.(\ref{K-symmetry1}) suggests that $H'$ fulfills the K symmetry:
\begin{equation}
 U^{-1}H'U = -H'^{\dagger}.
\end{equation}
Explicitly, $H'$ is given by
\[
H'=
\begin{pmatrix}
\frac{L-1}{2}F i&\frac{J}{2}&&&\\
\frac{J}{2}&\frac{L-3}{2}F i&\frac{J}{2}&&\\
&\ddots&\ddots&\ddots&\\
&&\frac{J}{2}&-\frac{L-3}{2}F i&\frac{J}{2}\\
&&&\frac{J}{2}&-\frac{L-1}{2}F i
\end{pmatrix},
\]
and the matrix element of transformation matrix $U$ is given by
\[
U_{ij}=(-1)^i \delta_{ij} .
\]
The existence of $K$ symmetry suggests that the eigenvalues of  Eq.(\ref{eq:Hamiltonian}) are either imaginary or distribute symmetrically about the imaginary axis (see Appendix \ref{sec:Gauge-transformation} for details).

Next we show that the model (\ref{eq:Hamiltonian}) fulfills a passive $\mathcal{PT}$ symmetry \cite{PT,passivePT,passivePT1}, i.e., the Hamiltonian
after making an overall energy shift fulfills
\begin{equation}
[\mathcal{PT}, \hat{H'}]=0, \label{eq:PT symmetry}
\end{equation}
where the space-reflection (parity) operator $\mathcal{P}$ and the time-reversal
operator $\mathcal{T}$ correspond to the following operations:
\begin{equation}
\mathcal{P} \hat{c}_{j} \mathcal{P} = \hat{c}_{L-j+1}, ~~~\mathcal{T} i \mathcal{T} = -i  .
\end{equation}
It is easy to check that Eq.(\ref{eq:PT symmetry}) is fulfilled due to the shifted on-site potential $V_{n}= i(\frac{L+1}{2}-n)F$ fulfilling $V_{n}=V_{L-n+1}^{*}$.
As usual, the eigenvalues of a $\mathcal{PT}$ symmetric system are either real or appear in conjugated pairs.  Consequently, the passive $\mathcal{PT}$ symmetry leads to the eigenvalues of Eq.(\ref{eq:Hamiltonian}) are either on the line of $E_{0}=-i\frac{L+1}{2}F$  or distribute symmetrically with respect to the line.


\begin{figure*}
\begin{centering}
\includegraphics[scale=0.35]{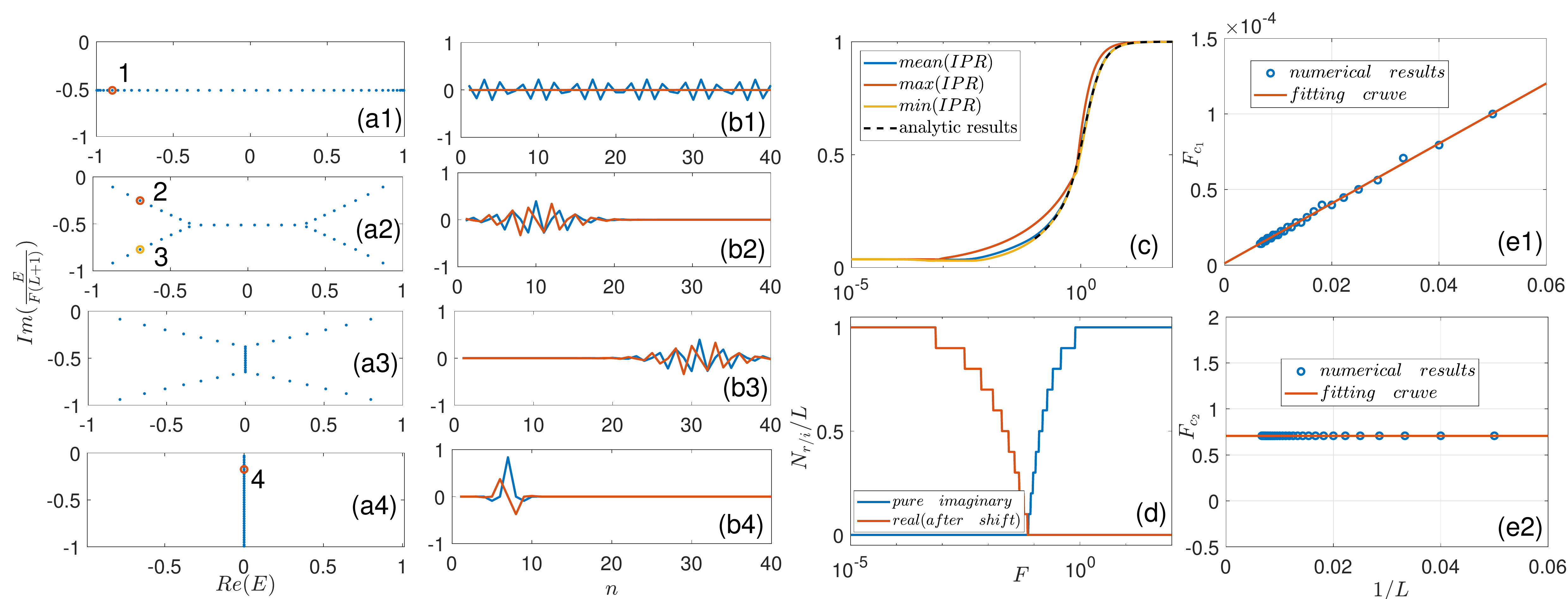}
\par\end{centering}
\caption{(a) Eigenvalues of Eq.(\ref{eq:Hamiltonian}) with different $F$: (a1)
$F=10^{-5}$; (a2) $F=0.05$; (a3) $F=0.1$; (a4) $F=1$. (b) Distribution of eigenstates corresponding
to the marked point in (a). The blue line represents the real part of the
eigenstate and the red line represent the imaginary part of the eigenstate.
(c) Mean IPR as a function of $F$. (d) Number of real eigenvalues
and pure imaginary eigenvalues (after the energy shift) as a function
of $F$. For (a)-(d), the system size $L=40$. (e) Finite-size analysis: (e1) Scaling behavior for  $F_{c_1}$. The fitting cruve satisfies: $F_{c_1}=0.001623\times L^{-0.9365}$. (e2) Scaling behavior for  $F_{c_2}$. The fitting cruve satisfies: $F_{c_2}=0.7079$.
\label{fig:changeF}}
\end{figure*}

When we increase the potential strength $F$, a $\mathcal{PT}$-symmetry breaking transition is expected to occur.
At the limit of $F=0$,
the Hamiltonian is Hermitian and all the eigenvalues are real. If we
add a very small $F$ which does not break the passive $\mathcal{PT}$ symmetry of $H'$, all the eigenvalues of $H$
have the same imaginary part $E_0=-i\frac{L+1}{2}F$.
When we increase $F$ over a threshold $F_{c_1}$, a $\mathcal{PT}$-symmetry
breaking occurs and the eigenvalues begin to deviate from $Im (E) = E_{0}$ at the
edge of the spectrum.
To see it clearly, we display the eigenvalues of the system with different strengths
of the imaginary potential $F$ in Fig.\ref{fig:changeF}(a).
When we continue to
increase $F$, more and more eigenvalues move to the imaginary axis. When $F$ exceeds another threshold $F_{c_2}$, all eigenvalues are distributed
on the imaginary axis as shown in Fig.\ref{fig:changeF}(a4), and the $K$ symmetry is restored with all eigenstates fulfilling the $K$ symmetry.

In Fig.\ref{fig:changeF}(b), we show some typical eigenvectors. When $F$
is small, all the eigenstates are extended in real space and fulfill the $\mathcal{PT}$ symmetry as shown in Fig.\ref{fig:changeF}(b1).
The $\mathcal{PT}$ symmetry broken
eigenstates corresponding to the conjugated pairs of eigenvalues marked in Fig.\ref{fig:changeF}(a2) are shown in Fig.\ref{fig:changeF}(b2) and Fig.\ref{fig:changeF}(b3), respectively. The eigenstate corresponding to the eigenvalue on the imaginary axis marked in Fig.\ref{fig:changeF}(a4) is shown in Fig.\ref{fig:changeF}(b4). This state is localized and fulfills the  $K$ symmetry. With the increase in $F$,  the eigenstates become more and more localized. To see it clearly, we plot the
mean IPR as a function of $F$ in Fig.\ref{fig:changeF}(c), where the mean
IPR is defined as $\langle IPR\rangle=\frac{1}{L}\sum_{m}\sum_{n}|\psi_{m}(n)|^{4}$, and $\psi_m(n)$ is the distribution of the $m_{th}$ wave function on the $n_{th}$ site.
We also plot the minimum and maximum values of IPR.
In order to determine the transition points between different regions,
we count the number of pure imaginary eigenvalues ($N_i$) and real eigenvalues ($N_r$)
after taking an energy shift $E'=E-E_{0}$ and display the fraction of real and imaginary eigenvalues in Fig.\ref{fig:changeF}(d).
From the figure, we can determine the $\mathcal{PT}$-symmetry
breaking point $F_{c_1}$, below which $N_r/L=1$, and the $K$-symmetry restoring point $F_{c_2}$, above which $N_i/L=1$.

When $L\rightarrow\infty$, we can get analytical expressions of eigenvalues and eigenstates of the Hamiltonian (\ref{eq:Hamiltonian}) {[}see Appendix \ref{sec:eigenvalues-and-eigenstates for large L}{]}. By solving $\hat{H}|\psi_{m}\rangle=E_{m}|\psi_{m}\rangle$
analytically, where $|\psi_{m}\rangle= \sum_n \psi_{m}(n) \hat{c}_n^{\dagger} |0\rangle $,  we get
\begin{equation}
E_{m}=-imF,m=1,2,3\dots\label{eq:limit-E}
\end{equation}
and
\begin{equation}
\psi_{m}(n)=c\mathcal{J}_{n-m}(-i\gamma)\label{eq:limit-EV}
\end{equation}
where $\gamma=J/F$, $c=1/\sqrt{\sum_{n}|\mathcal{J}_{n-m}(-i\gamma)|^{2}}$ is a normalization factor,
$E_{m}$ denotes the $m$-th eigenvalue, and $\psi_{m}$ is the corresponding eigenstate.
According to the property of the Bessel function, the eigenstate can be represented
by a modified Bessel function $I_{n}(x)$ and $\psi_{m}(n)=(i)^{m-n}cI_{n-m}(\gamma)$.
The eigenstate $\psi_{m}(n)$ is localized at the site $n=m$ and we find that the modulo square of the wavefunction $|\psi_{m}(n)|^2$ can be approximated by a Gaussian function \begin{equation}
|\psi_m(n)|^2 \approx a_0e^{-(\frac{n-m}{\sigma})^2 },
\end{equation}
where $a_0$ is a normalization constant and $\sigma$ is related to the localization length of the wavefunction by $$l_s=\sqrt{2} \sigma,$$ which fulfills $|\psi_m(m+l_s)|/|\psi_m(m)| = e^{-1}$.
For a given eigenstate, the wavefunction is mainly distributed in the region ${[}m-l_s,m+l_s {]}$.
The localization length decreases with the increase of $F$. Our numerical result indicates that  the localization length approximately fulfills a power-law relation $$l_s \propto \frac{1}{F^{\alpha}}$$ with $\alpha \approx 0.58$.
The eigenstate  $\psi_{m'}(n)$  can be achieved by shifting the localization center via a simple real-space translation. All eigenvalues are distributed
on the imaginary axis with the same level spacing.

\begin{figure}
\centering{}\includegraphics[scale=0.3]{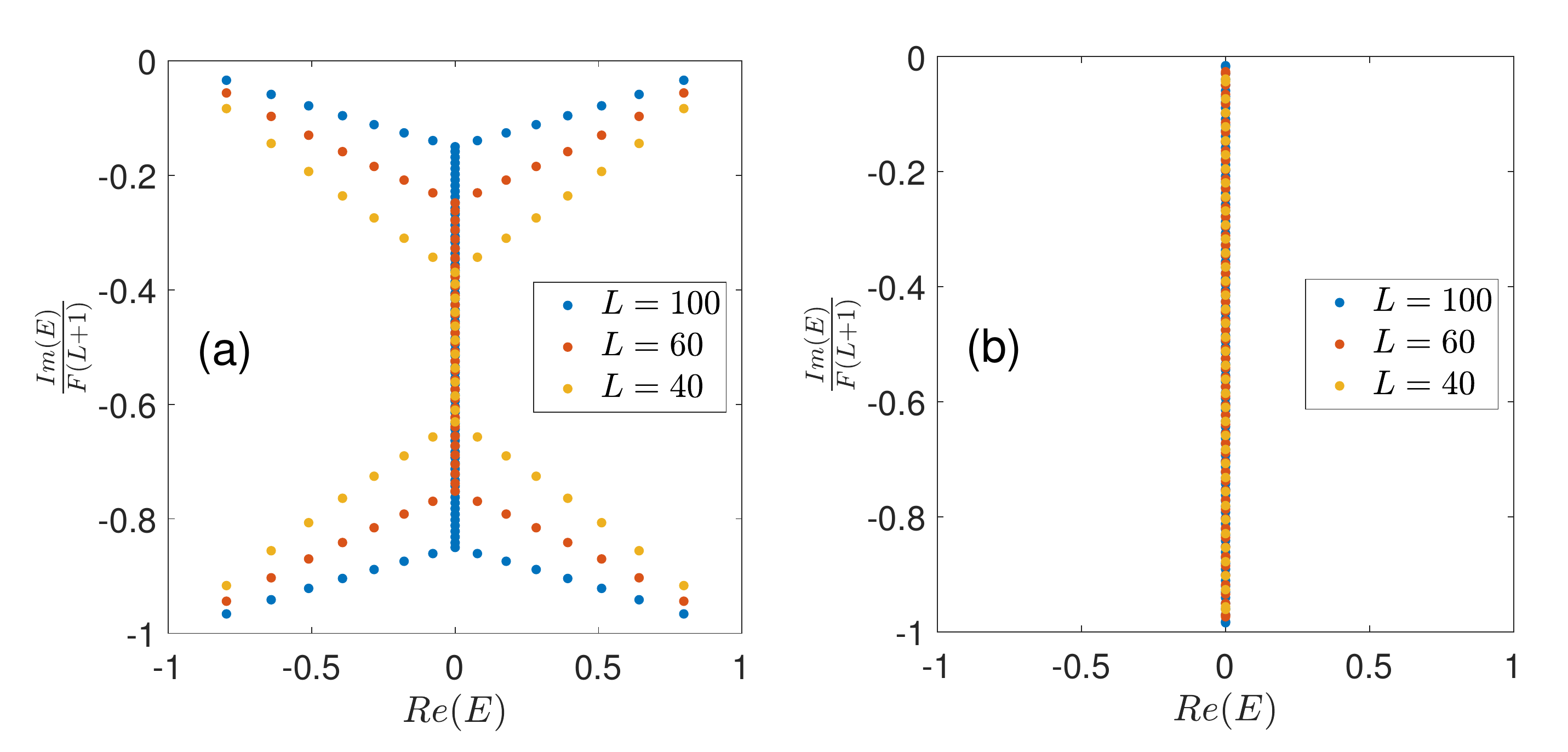}\caption{(a) Spectrum for $F=0.1$ with different system sizes. (b) Spectrum for $F=0.8$ with different system sizes. \label{fig:changeL}}
\end{figure}

Next, we consider the effects of the system size and the open boundary condition. We take finite-size analysis on the transition points. The numerical results are shown in Figs.\ref{fig:changeF}(e1) and (e2): $F_{c_1}$ shows a power-law decay when we increase the system size, whereas $F_{c_2}$ does not depend on the system size. The numerical results indicate that $F_{c_1}$ approaches zero and $F_{c_2} \approx  0.7079$ when $L$ tends to infinity.
Intuitively, $F_{c_2}$ can be roughly estimated from $l_{s} \approx 1$.
If the localization length $l_{s} < 1$, the boundaries do not affect the properties of localized state greatly, and the localized state of a finite size system can be still well described by the analytical expression (\ref{eq:limit-EV}). Here we emphasize that the analytical results given by Eqs.(\ref{eq:limit-E}) and (\ref{eq:limit-EV}) are valid description for the case with all eigenvalues on the imaginary axis, i.e., the case of $F>F_{c_2}$.
It follows that the IPR of states on the imaginary axis can be approximately expressed as
$I=\sum_{n}|c\mathcal{J}_{n-m}(-i\gamma)|^{4}$.
We display the IPRs in terms of the above analytic expression in Fig.\ref{fig:changeF}(c), which are shown to agree well with the numerical results.
When $l_{s}$ is much larger than $1$ and smaller than the system size, the boundaries have a significant impact on eigenstates close to the boundaries. For eigenstates far from the boundary, the impact of boundary for these eigenstates can be neglected.
These eigenstates distribute on the imaginary axis and can also be approximated by the analytic results (see Appendix \ref{sec:eigenvalues-and-eigenstates for large L}).
For eigenstates near the boundary, the impact of the boundary condition can not be neglected.
Consequently, there are always some eigenvalues
departing the imaginary axis, which distribute symmetrically with respect to the imaginary axis and the line of $E=E_0$.
As shown in Fig.\ref{fig:changeL}(a), for the system with a fixed $F=0.1$ and various sizes $L$, the numerical
results show that the numbers of states departing the imaginary axis are fixed for different system sizes.
Moreover, the real part of the corresponding eigenvalues are also size-independent and only determined by the strength
of the complex potential. When $F>F_{c_2}$, all the states distribute on the imaginary axis for systems with different sizes, as shown in Fig.\ref{fig:changeL}b.

\begin{figure}
\centering
\includegraphics[scale=0.302]{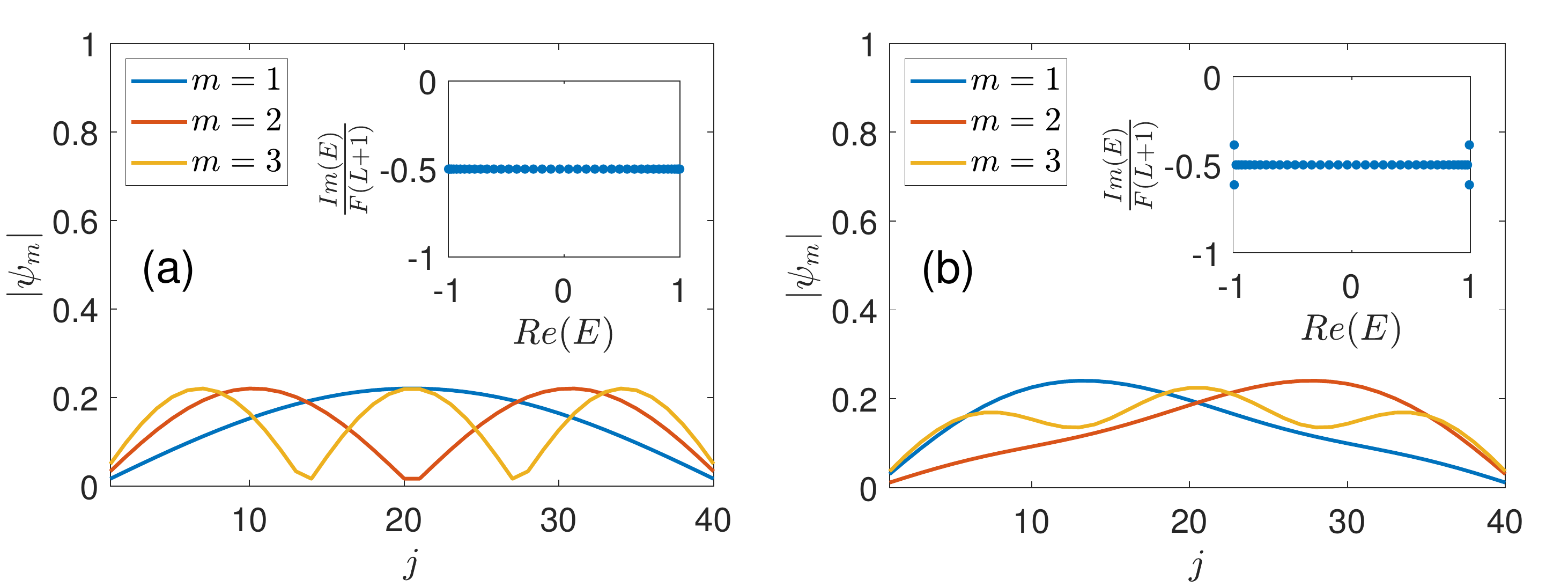}
\caption{ Absolute value of the eigenstates corresponding to the first three eigenvalues with the smallest real part before and after the passive $\mathcal{PT}$ symmetry breaking. We choose $L=40$ and the transition point is $F_{c_1}\approx 10^{-4}$.
So we choose the parameters as :(a) F=0.00001 (b) F=0.01.
Inset is the energy spectrum for different parameters.
We choose system size as $L=40$.
We use $m$ to label different eigenstates.
The sorting rules are as follows:
we first arrange the energy according to the real part from small to large, and those with the same real part are sorted according to the imag part from large to small.
\label{PTwavefunction}}
\end{figure}

Our numerical results have shown clearly that the transition point of passive $\mathcal{PT}$ symmetry breaking transition is size dependent and shall approach zero in the limit of $L\rightarrow \infty$. However, for a fixed $L$, there always exists a  $\mathcal{PT}$-unbroken region, in which all the eigenstates distribute on the whole lattice and are extended states. For $F>F_{c_1}$, the passive $\mathcal{PT}$ symmetry breaking transition happens. With further increase of $F$, stark-localization states occur, corresponding to states distributed on the imaginary axis. We note that the transition from the extended states to the stark-localization states is not a sharp transition.  The passive $\mathcal{PT}$ symmetry breaking transition does not occurs simultaneously with the localization transition.
To see it clearly, we display the absolute value of the eigenstates corresponding to the first three eigenvalues with the smallest real part before and after the passive $\mathcal{PT}$ symmetry breaking in Fig.\ref{PTwavefunction}.
In Fig.\ref{PTwavefunction}(a), we choose $F=0.00001$ which is smaller than $F_{c_1}$.
All the three eigenstates fulfill $\mathcal{PT}$ symmetry and are extended.
When we increase $F=0.01$ beyond $F_{c_1}$,  the first two eigenstates are $\mathcal{PT}$-symmetry broken and distribute asymmetrically about the center,
while the third eigenstate still preserves the $\mathcal{PT}$ symmetry,  as shown in Fig.\ref{PTwavefunction}(b).
However, the first two eigenstates are still extended. This indicates clearly that the $\mathcal{PT}$ symmetry breaking transition does not occur simultaneously with the localization transition.

\begin{figure}
\centering{}\includegraphics[scale=0.36]{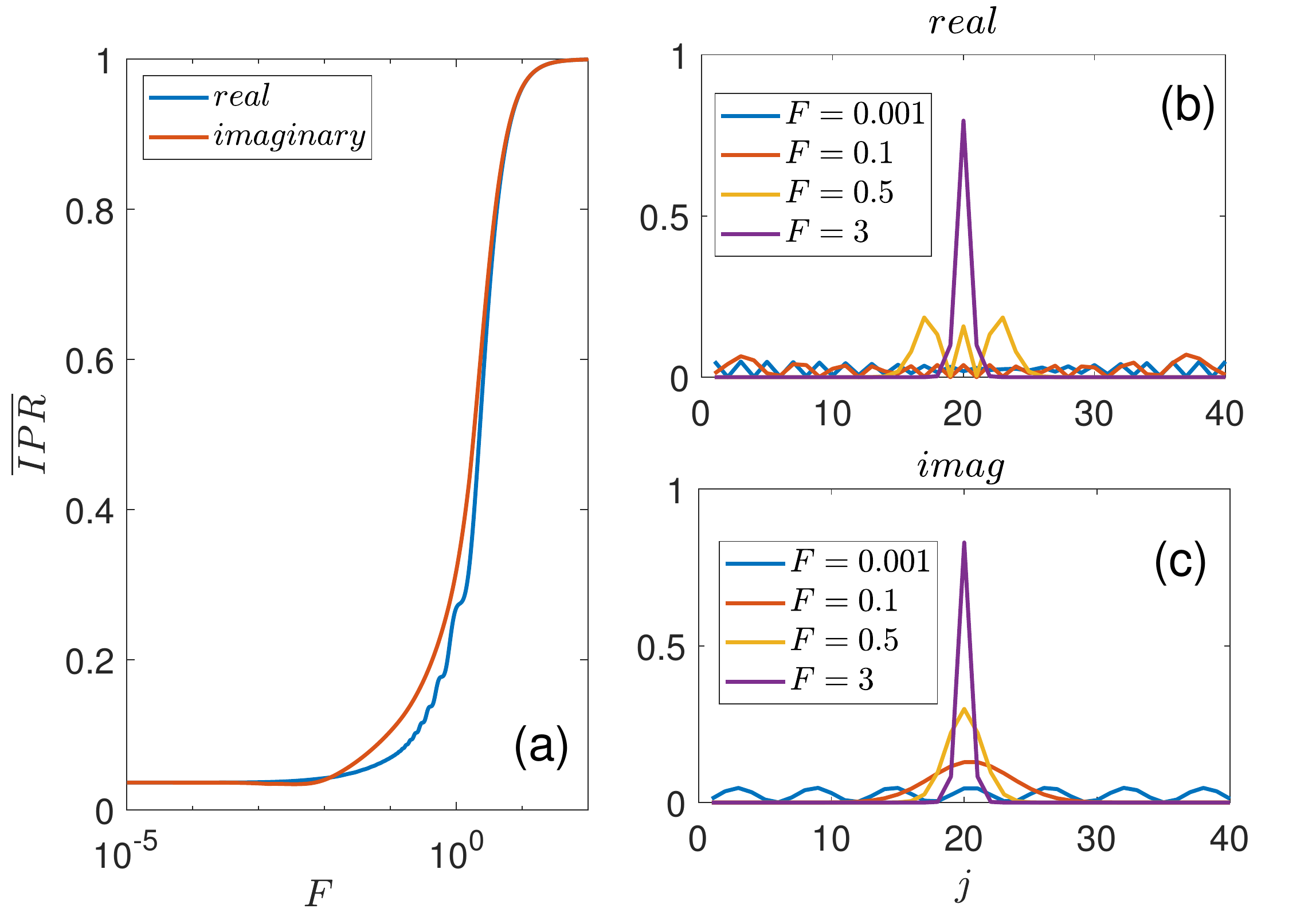}\caption{Comparsion between localization induced by real and imagnary potentials.
(a) Mean IPR with different $F$. (b) Distribution of $|\psi_{m}|^{2}$ with
real potential for different $F$. (c) distribution of $|\psi_{m}|^{2}$ with
imaginary potential for different $F$. We choose $m=21$ in the
figure. \label{fig:Comparsion-between-localization}}
\end{figure}

Before going to the next section, we would like to compare the localized eigenstates in real and imaginary stark ladder model. Explicitly, the stark ladder model is described by
\begin{equation}
\hat{H}=\frac{J}{2}\sum_{j=1}^{L-1}(\hat{c}^{\dagger}_{j+1}\hat{c}_j+h.c.)- F\sum_{j=1}^{L}j\hat{n}_j.
\end{equation}
It is known that the stark ladder model can also lead to localized eigenstates.
In Fig.\ref{fig:Comparsion-between-localization}, we display the mean IPR in Fig.\ref{fig:Comparsion-between-localization}(a) and the typical eigenstates with different $F$ in Fig.\ref{fig:Comparsion-between-localization}(b) and Fig.\ref{fig:Comparsion-between-localization}(c) for the real and imaginary stark ladder model, respectively. It is shown that the eigenstates in the imaginary stark ladder model is more localized than the real case and the shapes of eigenstates are also different.
Although both models exhibit stark localization, the physical meanings of the localization in real case and the imaginary case are different. For the real case, a linearly increasing potential suppresses the tunneling of the wavefunction and leads to localized eigenstates. As for the imaginary case, the imaginary potential leads to localized dissipation modes.

\section{Dynamical behaviors in the dissipative lattice\label{part:open system}}

The model of imaginary stark ladder can be realized in a lossy lattice with linearly increasing site-dependent dissipation. We consider the open quantum system with the evolution of density matrix
governed by the Lindblad master equation:
\begin{equation}
\frac{d\rho}{dt}=-i[\hat{H}_{0},\rho]-\sum_{\mu}(L_{\mu}^{\dagger}L_{\mu}\rho+\rho L_{\mu}^{\dagger}L_{\mu}-2L_{\mu}^{\dagger}\rho L_{\mu})=\mathcal{L}\rho, \label{Lindbald}
\end{equation}
where $\rho$ is the density matrix,
$$
\hat{H}_{0}=\frac{J}{2}\sum_{j}(\hat{c}_{j+1}^{\dagger}\hat{c}_{j}+\hat{c}_{j}^{\dagger}\hat{c}_{j+1})
$$
is the Hamiltonian and $L_{\mu}$ are the Lindblad operators describing
the dissipative process. The index $\mu$ denotes the
dissipation channel. Here we consider the dissipation operators acting on
each lattice site, i.e, $\mu=j$ with $j=1, \cdots, N$, and taking the form of
\begin{equation}
L_{j}= \sqrt{\gamma_j} \hat{c}_{j}= \sqrt{jF}\hat{c}_{j} , \label{Lj}
\end{equation}
where $\hat{c}_{j}$ is the fermionic annihilation
operator  and $\gamma_{j}=jF$ denotes the site-dependent dissipation strength, which increases linearly with the site $j$.

Define
\begin{equation}
\hat{H}_{eff}=\hat{H}_{0}-i L_{\mu}^{\dagger}L_{\mu},
\end{equation}
and then Eq.(\ref{Lindbald}) can be rewritten as
\begin{equation}
\frac{d\rho}{dt}=-i(\hat{H}_{eff}\rho-\rho \hat{H}_{eff}^{\dagger})+2\sum_{\mu}L_{\mu}^{\dagger}\rho L_{\mu},
\end{equation}
where $\hat{H}_{eff}$ is the effective non-Hermitian Hamiltonian and the other terms $2\sum_{\mu} L_{\mu}^{\dagger}\rho L_{\mu}$ are called jump terms.
For the case with $L_{\mu}$ given by Eq.(\ref{Lj}), the effective non-Hermitian Hamiltonian has the same form as Eq.(\ref{eq:Hamiltonian}). When the jump terms are omitted, the dynamics is governed by the effective non-Hermitian Hamiltonian.
In general, the jump terms also contribute to the dynamical behaviors and non-Hermitian dynamics is only a short time approximation of Lindblad evolution.
However, for the case with the initial state in the subspace of $N=1$, the dynamics of the single particle correlation function governed by the Lindblad equation and non-Hermitian effective Hamiltonian are equivalent (more details about the connection and difference between them can be found in Appendix \ref{LindbladVsNonHermitian}).

In the scheme of Lindblad equation, the time evolution of the single-particle correlation
$\Delta_{jj'}(t)=\text{Tr}[\hat{c}_{j}^{\dagger}\hat{c}_{j'}\rho(t)]$ can be determined
by the damping matrix \cite{chiraldamping}
\begin{equation}
X=ih^{T}-M^{T},
\end{equation}
where $\left(h\right)_{jk}=J(\delta_{j,k+1}+\delta_{j+1,k})/2$,
$M_{jk}=\gamma_j \delta_{jk}$ and the dynamics
of the single-particle correlation is governed by
\begin{equation}
\frac{d\Delta(t)}{dt}=X\Delta(t)+\Delta(t)X^{\dagger}.
\end{equation}
We note that the $X$ matrix can be written as $X=iH^{*}$, where $H$
is the matrix form of the Hamiltonian (\ref{eq:Hamiltonian}).

\begin{figure}
\centering
\includegraphics[scale=0.38]{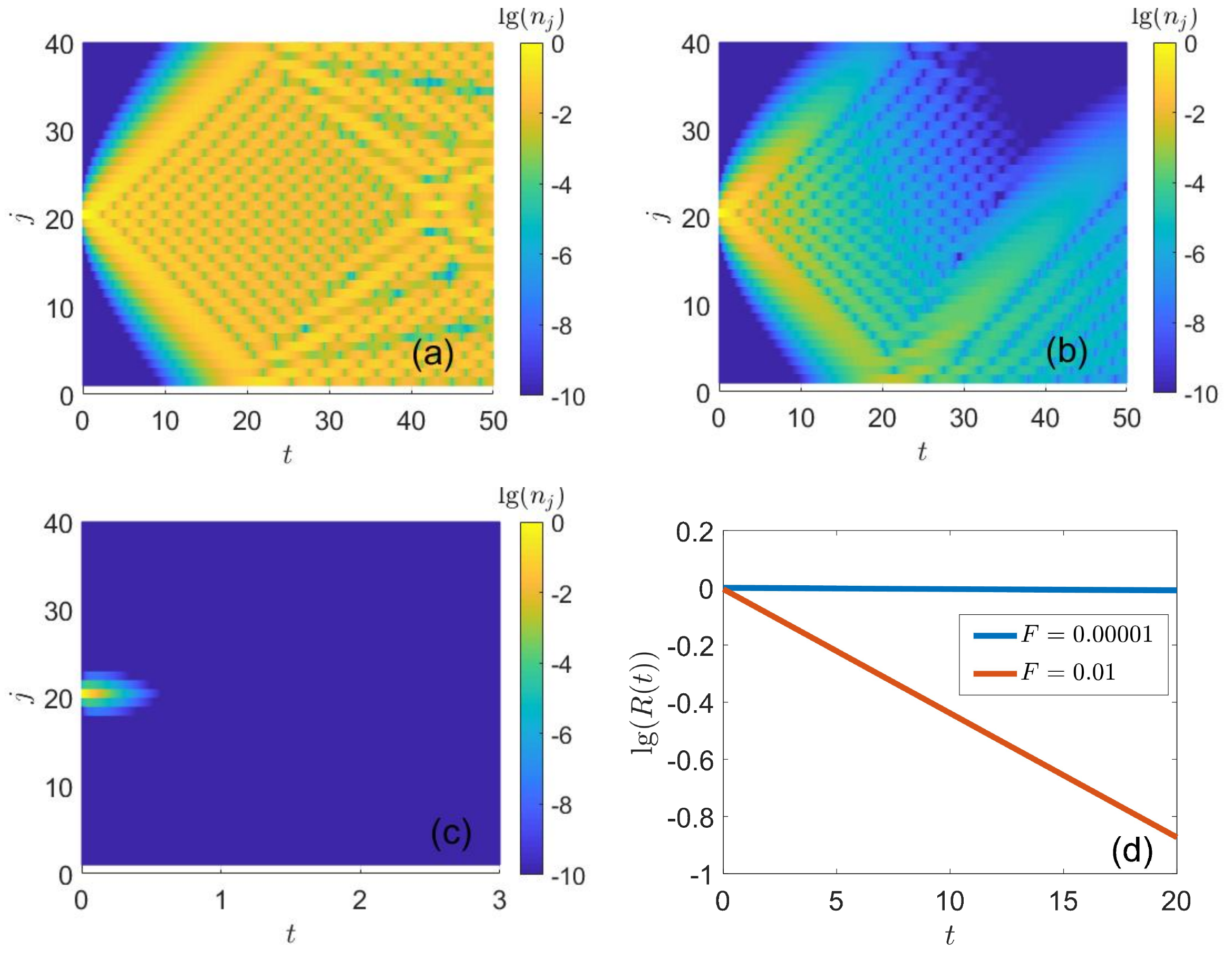}
\caption{
(a-c) Non-hermitian wavepacket dynamics for different $F$: (a) $F=0.00001$, (b) $F=0.01$, (c) $F=1$.
(d) $R(t)=n_{25}(t)/n_{15}(t)$ for different $F$.
We choose $L=40$ in our calculation.
\label{fig:NDF}}
\end{figure}

First, we study the time evolution of the wavepacket.
In Figs.\ref{fig:NDF}(a)-(c), we display the time evolution of the particle density distribution $n_j(t)$ ($n_j(t)=\Delta_{jj}(t)$) for various $F$ with the initial state chosen as a wavepacket localized in the center of the lattice.
When $F=0.0001$, the passive $\mathcal{PT}$ symmetry of the corresponding imaginary stark ladder model is unbroken, and the time evolution of wavepacket is very similar to a free fermion system but with a global decaying rate.
When we increase $F$ over $F_{c_1}$, for example $F=0.01$, the $\mathcal{PT}$ symmetry is broken.  The density distribution displays obviously different decaying behaviors on the two sides.
When $F$ is large enough, e.g., $F=1$, the evolution of wavepacket displays a localized damping behavior. As shown in Fig.\ref{fig:NDF}c, the wavepacket decays very quickly on its initial position and the spreading to another sites are suppressed due to the emergence of stark-localization modes.
In Fig.\ref{fig:NDF}(d), we display the ratio of density on the 25th site and 15th site $R(t)=\frac{n_{25}(t)}{n_{15}(t)}$ for $F=0.00001$ and $0.01$.
It is shown
that $R(t) \approx 1$ for $F=0.00001$, whereas $R(t)\propto \exp(-\beta t)$ for $F=0.01$, where $\beta$ is a constant. Although $R(t)$ exhibits obviously different behaviors for $F=0.00001$  and $0.01$, we note that $R(t)$ is not a suitable indicator for distinguishing the passive $\mathcal{PT}$ symmetry breaking and unbroken regions. No sharp change around $F_{c_1}$ can be observed.

\begin{figure}
\centering
\includegraphics[scale=0.36]{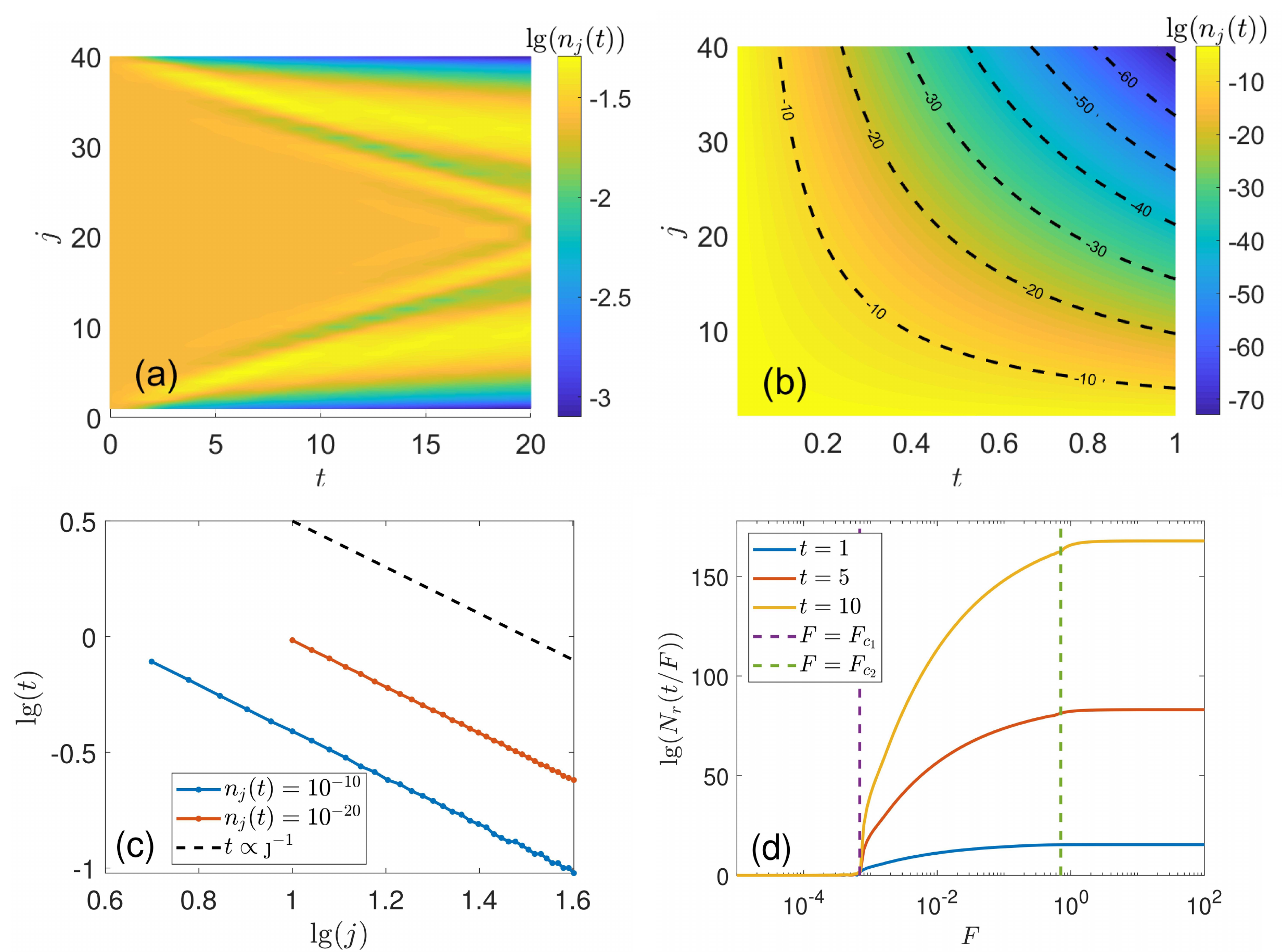}
\caption{(a)-(b)Time evolution of density distribution for system with $L=40$ and different strengths of
the potential: (a)$F=0.00001$. (b)$F=1$. The dashed lines denote contours, along which the local densities are identical.
(c) We plot $\lg(t)$ versus $\lg(j)$ for contours in (b), which show $t$ and $j$ fulfilling $t\propto j^{-1}$. (d) We plot $N_r(t/F)$ versus $F$ for various time $t$. $N_r(t/F)$ exhibits different behaviors in different regions.
\label{fig:dynamical_changeF}}
\end{figure}

Then we choose the initial state as a single excitation, which is uniformly distributed on the lattice: $|\psi(0)\rangle=\sum_j \frac{1}{\sqrt{L}}|j\rangle$,  and show the dynamical behavior with different dissipation strengths.
In Fig.\ref{fig:dynamical_changeF}(a-b), we plot the time evolution for $F=0.00001$ and $F=1$ and they display very different behaviors.
We display the time evolution of the density distribution $n_j(t)$ for the system with $F=1$ in Fig.\ref{fig:dynamical_changeF}(b) and label contours of the density distribution by using the dashed lines.
It is shown that the density distribution at different sites decay in different decaying rates.  By analyzing the data of contours, we find that the contours can be well fitted by hyperbolic curves described by $j\propto \frac{1}{t}$ as shown in Fig.\ref{fig:dynamical_changeF}(c).
This behavior can be explained by the existence of localized modes and equispaced energy spectrum in the region of $F>F_{c_2}$, which gives rise to
\begin{equation}
n_j(t) \approx \frac{1}{L}e^{-2Fjt} \label{njt}
\end{equation}
(see Appendix \ref{open dynamical} for more details).

In order to characterize the difference of dynamical behavior in passive $\mathcal{PT}$ symmetry breaking and unbroken region, we define a rescaled particle number: \begin{equation}
N_r(t)=\exp[F(L+1)t] \sum_{j=1}^{L} n_j(t). \label{Nrt}
\end{equation}
In the passive $\mathcal{PT}$ symmetry unbroken region, $N_r(t)=1$.
While in the passive $\mathcal{PT}$ symmetry breaking region, $N_r(t)$ shows an exponential increase.
We plot the rescaled particle number versus the dissipation strength $F$ for various rescaled time ${t}/{F}$.
The results are shown in Fig.\ref{fig:dynamical_changeF}(d), which indicates clearly the occurrence of a sharp change of $N_r$ near the passive $\mathcal{PT}$ breaking transition point $F_{c_1}$\cite{NH_LME}. Substituting Eq.(\ref{njt}) into Eq.(\ref{Nrt}), we get $$N_r(t/F)=\frac{1}{L} \frac{\sinh(Lt)}{\sinh(t)}$$ for $F>F_{c_2}$. It implies that $N_r$ shall not change with $F$ when $F>F_{c_2}$, consistent with our numerical results in Fig.\ref{fig:dynamical_changeF}(d).

\begin{figure}
\centering
\includegraphics[scale=0.3]{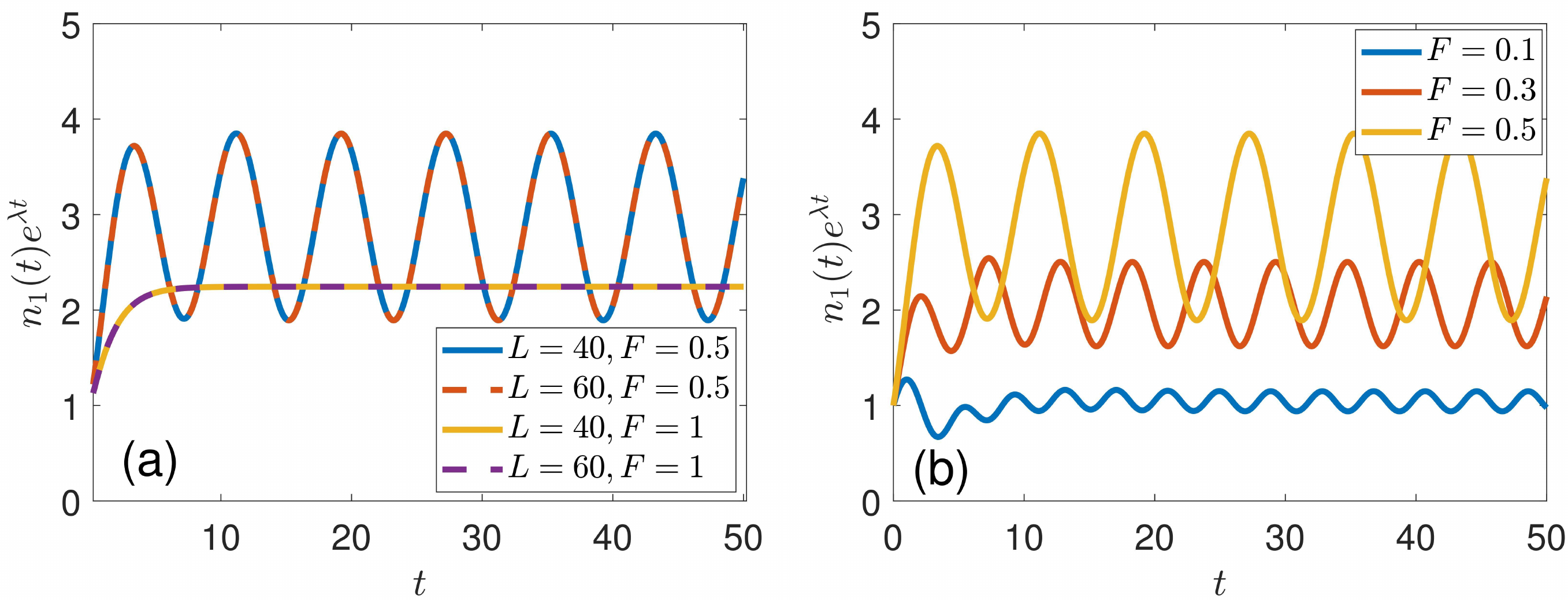}
\caption{ Time evolution of a rescaled local density distribution
at the first site. We choose $L=40,60$ and $F=0.5,1$ in (a)
and  $L=40$ and $F=0.1,0.3,0.5$ in (b). \label{decay}}
\end{figure}

When $F<F_{c_2}$, the $K$ symmetry is broken and complex eigenvalues appear in the symmetry breaking region. The real part of these complex eigenvalues are size-independent (see Fig.\ref{fig:changeL}(a)) and shall lead to dynamical oscillating behavior, which can be detected by the size-independent oscillation in the boundary.
In Fig.\ref{decay}, we display the time evolution of a rescaled density distribution
at the first site. Here a rescaled factor $e^{\Lambda t}$ is introduced to eliminate the decaying behavior caused by the imaginary
part of the complex eigenvalue, where $\Lambda$ is the average speed of decay.
From Fig.\ref{decay}(a), we can see that the boundary oscillating behavior is insensitive to the system size for $F=0.5$, whereas no oscillation is observed for $F=1$.
Fig.\ref{decay}(d) shows the frequency of oscillation decreases with the increase of $F$ and this behavior is consistent with our analysis of the spectrum: when we increase the potential strength, the real part of the complex eigenvalue becomes smaller, corresponding to a larger period.%

\begin{figure}[H]
\centering{}
\includegraphics[scale=0.3]{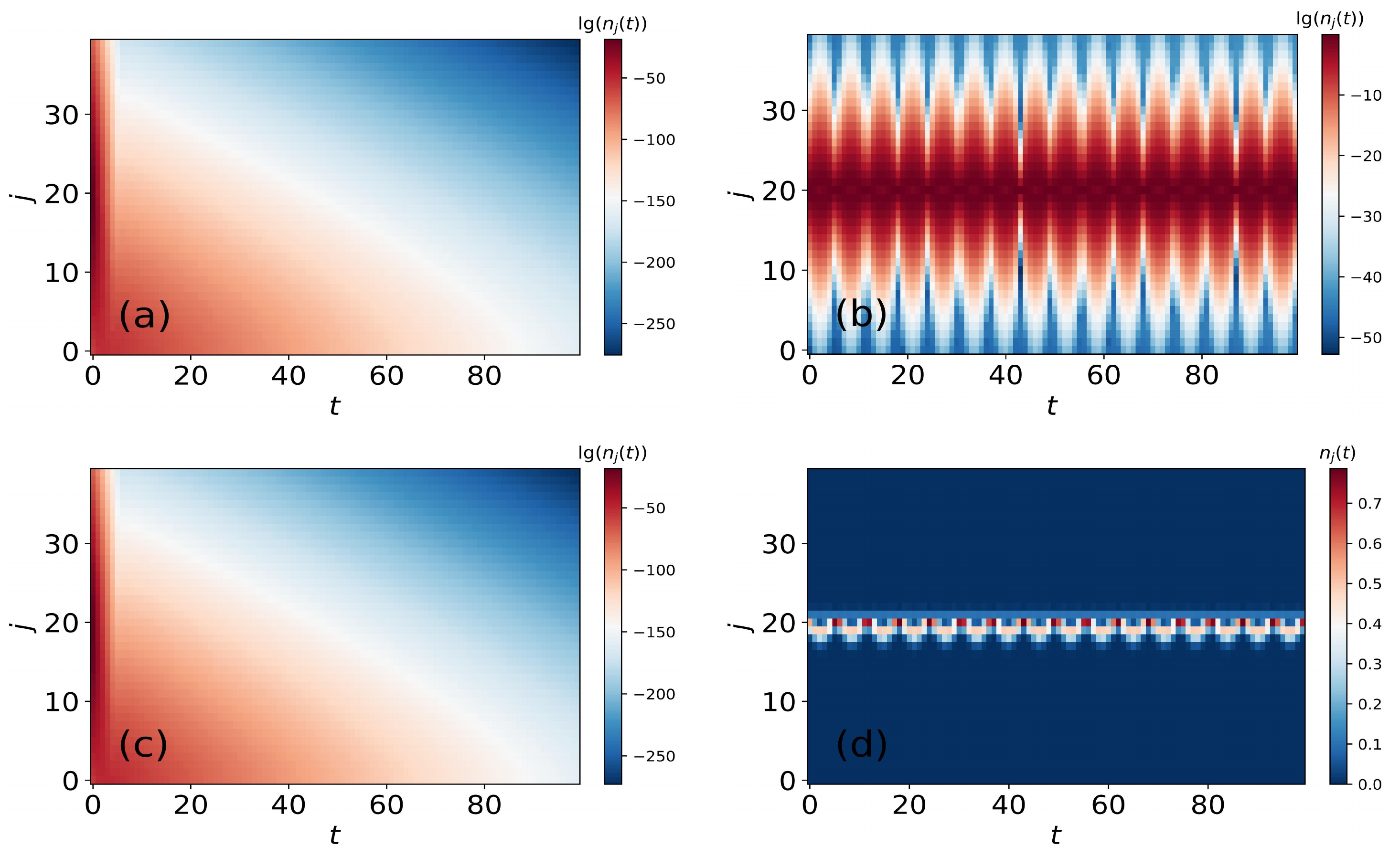}
\caption{(a)(c) Time evolution of $n_j(t)$ in the imaginary stark ladder model. (b)(d) Time evolution of $n_j(t)$ in the  stark ladder model. In (a-b), we choose $L=40$ , $F=1$ and the initial state is chosen as the state localized at the center site of the system. In (c-d), we choose $L=40$ , $F=1$ and the initial state is chosen as a Gaussian wavepacket ($|\psi \rangle=c_0\sum_j \exp(-\beta(j-L/2)^2) |j\rangle$, $c_0$ is the normalization factor) localized at the center of the system. We choose $\beta=1$ in our results.   \label{compare_bloch_oscillation}}
\end{figure}

Finally, we compare the time evolution between the real and imaginary stark ladder models.
It is well known that the real stark ladder model can exhibit unique dynamical behaviors:
when the initial state is localized at one site, the dynamical behavior exhibits breathe modes, whereas it
exhibits Bloch oscillation if the initial state is chosen as a Gaussian wavepacket.
A natural question is the fate of the breathe mode and Bloch oscillation in the imaginary stark ladder model.
In Fig.\ref{compare_bloch_oscillation}. we display the time evolution of the density distribution in different models with different initial states.
In Fig.\ref{compare_bloch_oscillation}(a) and Fig.\ref{compare_bloch_oscillation}(c), we plot the time evolution of $n_j(t)$ in the imaginary stark ladder model with the initial state localized at one site and given by a Gaussian wavepacket, respectively.
The time evolution shows similar results for different initial states in the imaginary stark ladder model.
As a comparison, we show the breathe mode and Bloch oscillation in Fig.\ref{compare_bloch_oscillation}(b) and Fig.\ref{compare_bloch_oscillation}(d) with the initial state localized at one site and given by a Gaussian wavepacket, respectively.
Our numerical results show that both breathe mode and Bloch oscillation disappear in the imaginary stark ladder model.


\section{Summary \label{part:Conclusion-and-discussion}}
In summary, we propose and study the imaginary stark ladder model and its realization in a lossy lattice with linearly increasing on-site dissipation.
We demonstrate that the model possesses a passive $\mathcal{PT}$ symmetry and $K$ symmetry and increasing the strength of the imaginary potential can lead to $\mathcal{PT}$-symmetry breaking and $K$-symmetry restoration. We determine the transition points numerically and make finite-size analysis, which indicates the  $\mathcal{PT}$-symmetry breaking point $F_{c_1}$ approaches zero in the large size limit and the $K$-symmetry restoring point $F_{c_2} \approx 0.7079$ is size independent. When the dissipation strength is above $F_{c_2}$, all the eigenstates are localized and can be well described by the analytic result obtained in the limit $L\rightarrow \infty$.
By studying the lattice with linearly increasing on-site dissipation described by Lindblad equation,  we show that the dynamical evolution of density distribution displays distinct behaviors in different regions of the corresponding non-Hermitian Hamiltonian.

\begin{acknowledgments}
We thank Zhenyu Zheng, Xueliang Wang and Chun-hui Liu for helpful discussions.
The work is supported by the NSFC under Grants No.12174436, No.11974413
and No.T2121001, National Key
Research and Development Program of China (Grant No.
2021YFA1402104), and the Strategic Priority Research Program of Chinese Academy of Sciences under Grant No. XDB33000000.
\end{acknowledgments}

\appendix

\section{The $K$ symmetry \label{sec:Gauge-transformation}}

Given the transformation: $\hat{U}=\Pi_{j=1}^N [\hat{c}_j^\dagger + (-1)^j\hat{c}_j] $, it is easy to check
\begin{align*}
\hat{U}^{-1} \hat{c}_j \hat{U} &= (-1)^{j}\hat{c}^\dagger_j, \\
\hat{U}^{-1} \hat{c}^\dagger_j \hat{U} &= (-1)^{j}\hat{c}_j.
\end{align*}
Then we can prove that $\hat{H'}$ fulfills the $K$ symmetry under the transformation $\hat{U}$:
\begin{align*}
\hat{U}^{-1}\hat{H'}\hat{U} & =-\frac{J}{2}\sum_{j}(\hat{c}_j\hat{c}_{j+1}^{\dagger} +\hat{c}_{j+1}\hat{c}_{j}^{\dagger} )\\
 & -iF\sum_{j}(j -\frac{L+1}{2})\hat{c}_j \hat{c}_j^{\dagger}\\
= & \frac{J}{2}\sum_{j}(\hat{c}^{\dagger}_j\hat{c}_{j+1} +\hat{c}^{\dagger}_{j+1}\hat{c}_{j})\\
 & +iF\sum_{j}(j -\frac{L+1}{2})\hat{c}_j^{\dagger} \hat{c}_j\\
 & -iF\sum_{j}(j -\frac{L+1}{2})\\
= & \hat{H'}^{\dagger} .
\end{align*}
Since
$
\hat{H'}=\textbf{c}^\dagger H' \textbf{c},
$
from $\hat{U}^{-1}\hat{H'}\hat{U}=\hat{H'}^\dagger$, we get
\begin{align*}
\hat{U}^{-1}\hat{c}^\dagger_m H'_{mn} \hat{c}_n\hat{U}&=\hat{c}^\dagger_m {H'}^{*}_{nm} \hat{c}_n,\\
\hat{U}^{-1}\hat{c}^\dagger_m \hat{U}\hat{U}^{-1} H'_{mn} \hat{c}_n\hat{U}&=\hat{c}^\dagger_m {H'}^{*}_{nm} \hat{c}_n,\\
U_{mm'}\hat{c}_{m'}H'_{mn} U_{nn'}\hat{c}_{n'}^\dagger  &=\hat{c}^\dagger_m {H'}^{*}_{nm} \hat{c}_n,\\
-U_{mm'} H'_{mn} U_{nn'} \hat{c}_{n'}^\dagger \hat{c}_{m'}  &=\hat{c}^\dagger_{n'} {H'}^{*}_{m'n'} \hat{c}_{m'},
\end{align*}
where $H'_{nm}$ represents the element of $H'$. Then it follows
$ (U^{-1}H'U)^T = -H'^*$,
which is equivalent to
\[
U^{-1}H'U = -H'^{\dagger}.
\]

The existence of $K$ symmetry implies that if $E_m$ is an eigenvalue of $H'$, then $E_{m'}=-E_m^*$ is also an eigenvalue of $H'$. This can be proved straightforwardly: Given
$H' \psi_m =E_m \psi_m$,
where $\psi_m$ is the $m_{th}$ eigenvector of the matrix $H'$.
Then we can construct a state given by $\psi'_{m}=U \psi_{m}^L$ where $(\psi^L_m)^\dagger H'=E_m (\psi^L_m)^\dagger$.
It follows
\begin{align*}
H' \psi'_{m} & =H'U \psi^L_m\\
= & -UH'^{\dagger}\psi^L_m\\
= & -E_m^{*}U\psi^L_m \\
= & -E_m^{*} \psi'_{m},
\end{align*}
i.e., $-E_m^{*}$ is also an eigenvalue of $H'$ with $\psi'_{m}$ being an eigenvector of $H'$. This means that the eigenvalues of $H'$ are either pure imaginary or complex pairs distributing symmetrically about the imaginary axis.
The Hamiltonian before and after the energy shift operation only differs by an imaginary constant.
Thus $E_{m'}=-E_m^*$ still holds true for the energy spectrum of the Hamiltonian without energy shift.

\begin{figure}
\begin{centering}
\includegraphics[scale=0.2]{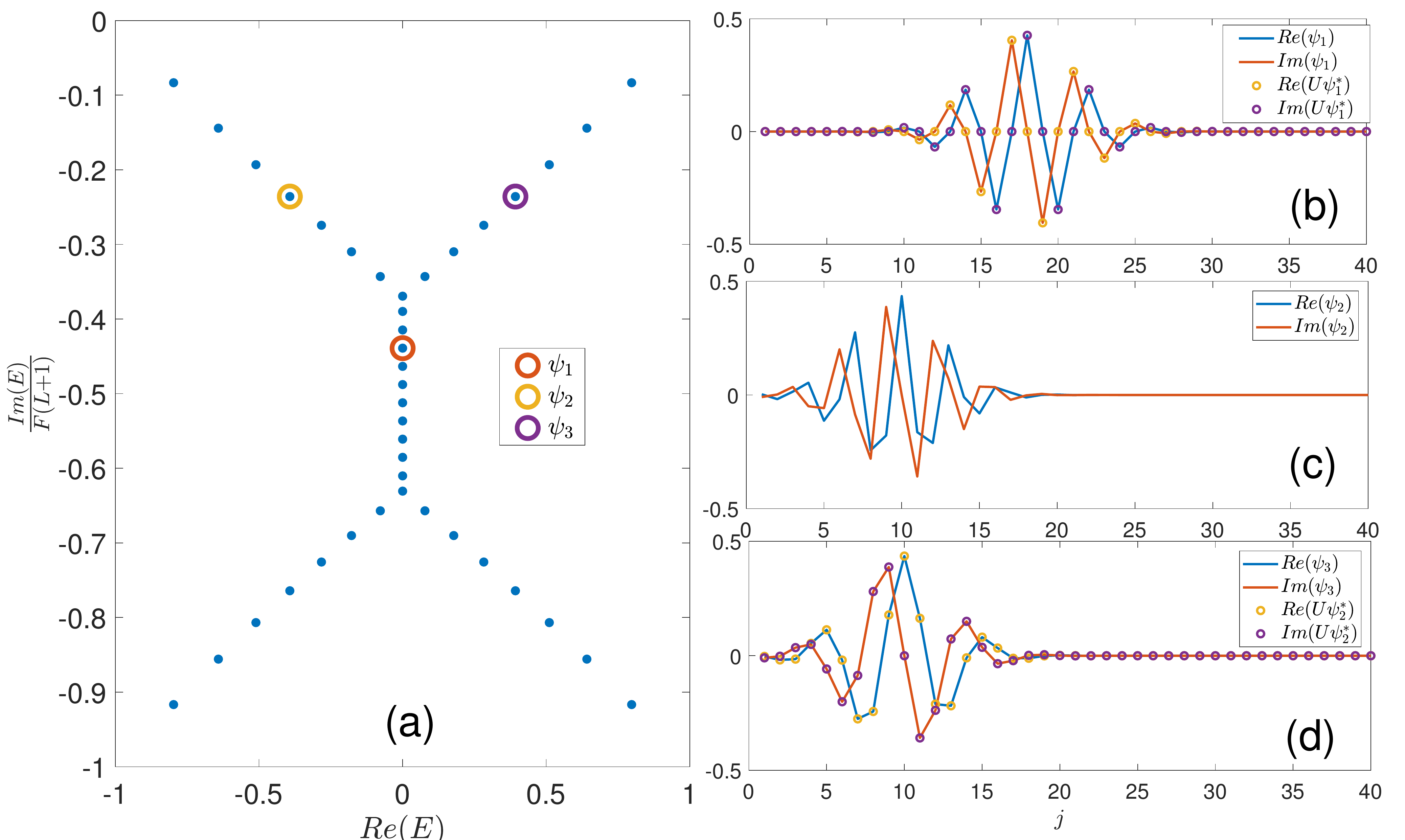}
\par\end{centering}
\caption{Numerical evidence for the $K$-symmetry broken. (a) Spectrum for $F=0.1$. (b-d) Distribution of eigenstates corresponding to the marked point in (a). (b) A typical eigenstate with unbroken $K$-symmetry. The eigenstate keeps invariant under the transformation of $\tilde{U}=UK$. (c-d) Typical eigenstates with broken $K$-symmetry. The eigenstates can be connected by the transformation  of $\tilde{U}$: $\psi_3= UK \psi_2 =U \psi_2 ^*$.\label{fig:Ksymmetry}}
\end{figure}

In Fig.\ref{fig:Ksymmetry}, we display the spectrum for the system with $F=0.1$ and several eigenstates.
In our model, $\psi^L_m=\psi^*_m$.
If an eigenstate fulfills $K$ symmetry, it is invariant under the transformation of $\tilde{U}=UK$, where $K$ denotes the complex conjugation, i.e., it fulfills $$\psi_{m}=\tilde{U}\psi_{m} =U \psi_{m}^{*}.$$
Our numerical result shows that the eigenstate, such as $\psi_{1}$, indeed fulfills $\psi_{1}=U \psi_{1}^{*}$. Then we check
the eigenstates $\psi_{2},\psi_{3}$, whose eigenvalues distribute symmetrically about the imaginary axis,
can be connected by $\psi_{3}= \tilde{U}\psi_{2}= U\psi_{2}^{*}$. The numerical results indicate that the eigenstates with the same imaginary part can be connected
by $\psi_{3}=U \psi_{2}^{*}$.
Here, we note that the subscripts $m=1,2,3$ just correspond to states labeled in Fig.\ref{fig:Ksymmetry}(a).

\section{Eigenvalues and eigenstates for $L\rightarrow\infty$\label{sec:eigenvalues-and-eigenstates for large L}}
In this appendix, we derive the analytical solution for the imaginary stark ladder in the limit of  $L\rightarrow\infty$ by following a method similar to the Ref.\cite{stark-localization-solve}.
After inserting a set of orthonormal basis vectors $\mathbb{I}=\sum_n|n\rangle\langle n|$, the Hamiltonian can be written as:
\begin{equation*}
\hat{H} =\frac{J}{2}\sum_{n}(|n+1\rangle\langle n|+|n\rangle\langle n+1|) -i F\sum_{n}n|n\rangle\langle n| .
\end{equation*}

Introducing a Fourier transform:
\[
|k\rangle=\sum_{n}\sqrt{\frac{1}{2\pi}}e^{ink}|n\rangle,
\]
then we can get the matrix element
\begin{align*}
\langle k'|\hat{H}|k\rangle & =\frac{J}{2}\sum_{n}(\langle k'|n\rangle\langle n+1|k\rangle+\langle k'|n+1\rangle\langle n|k\rangle)\\
 & -iF\sum_{n}\langle k'|n\rangle\langle n|k\rangle n\\
= & \frac{J}{2}\sum_{n}(\frac{1}{2\pi}e^{-ink'}e^{i(n+1)k}+\frac{1}{2\pi}e^{-i(n+1)k'}e^{ink})\\
 & -iF\sum_{n}e^{-ink'}e^{-inkd}\frac{n}{2\pi}\\
= & J\delta(k'-k)\cos(k)+F\delta(k'-k)\frac{d}{dk} .
\end{align*}
Since $\langle k'|\hat{H}|k\rangle=\delta(k'-k)H(k)$, we can get
\[
H(k)=J\cos(k)+F\frac{d}{dk}.
\]
Next, we solve the equation
\begin{equation}
E\psi(k)  =J\cos(k)\psi(k)+F\frac{d}{dk}\psi(k),
\end{equation}
or equivalently,
\begin{equation}
\frac{d}{dk}\psi(k)  =\frac{E}{F}\psi(k)-\frac{J}{F}\cos(k)\psi(k).
\end{equation}
Then it follows
\begin{align*}
\psi(k) & =e^{\int\frac{E}{F}-\frac{J}{F}\cos(k)dk}\psi(0)\\
 & =e^{\frac{E}{F}k-\frac{J}{F}\sin(k)}\psi(0).
\end{align*}
Besides, we have $\psi(k+2\pi)=\psi(k)$, which gives rise to
\begin{align*}
E_{m} & =-imF,m=0,\pm1,\pm2,\dots\\
\psi_{m}(k) & =e^{\frac{E_{m}}{F}k+\frac{J}{F}\sin(k)}\psi(0)=e^{-ikm-\frac{J}{F}\sin(k)}\psi(0) .
\end{align*}
Choosing a normalization factor $c$, then we change the wavefunction
into real space:
\begin{align*}
\psi_{m}(n) & =c\int_{-\pi}^{\pi}dk\langle n|k\rangle\langle k|\psi_{m}\rangle\frac{1}{2\pi}\\
= & c\frac{1}{2\pi}\int_{-\pi}^{\pi}dke^{ikn}e^{-ikm-\frac{J}{F}\sin(k)}\\
= & c\frac{1}{2\pi}\int_{-\pi}^{\pi}dke^{ik(n-m)-\frac{J}{F}\sin(k)} .
\end{align*}
Defining $J/F=\gamma$ and making use of the definition of Bessel function
$\mathcal{J}_{\alpha}(x)=\frac{1}{2\pi}\int_{-\pi}^{\pi}d\tau e^{i(\alpha\tau-x\sin(\tau))}$, we have
\[
\psi_{m}(n)=c\mathcal{J}_{n-m}(-i\gamma),
\]
where $m=1,2,\dots n$ in our model.

\begin{figure}
\centering
\includegraphics[scale=0.3]{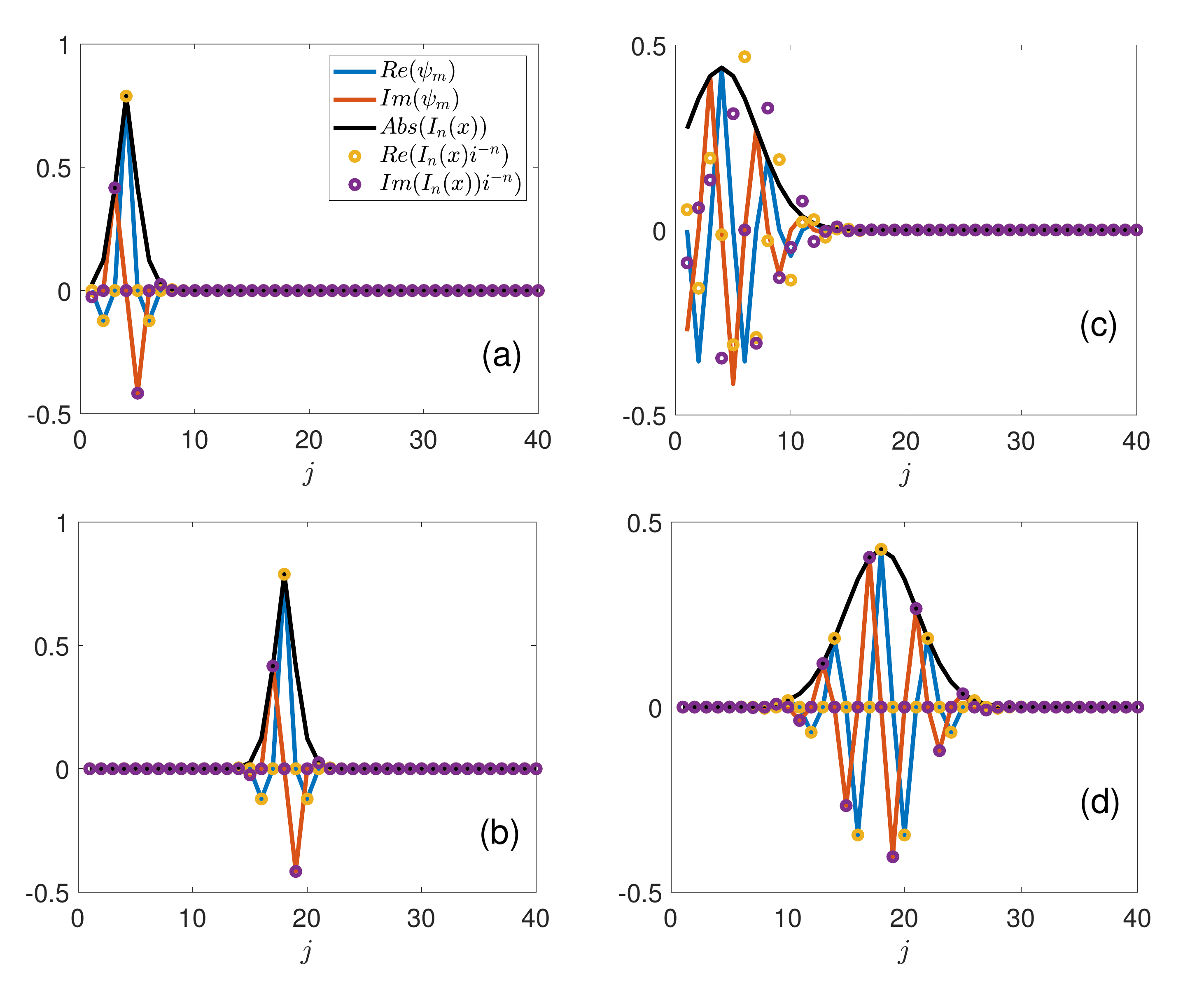}
\caption{ Comparison between numerical and analytical results for typical eigenstates.
The parameters in figure are chosen as:
(a) $F=0.8, m=7$.
(b) $F=0.8, m=18$.
(c) $F=0.1, m=7$.
(d) $F=0.1, m=18$.
We choose $L=40$.
\label{fig:Q2}}
\end{figure}

Next we show more details about the eigenstates.
There is a relationship that
\[
J_{n}(ix)=i^{-n}I_{n}(x)
\]
when $x$ is real and $I_{n}(x)$ is Modified Bessel function\cite{handbook of math}.
That means the eigenstate can be described by $I_{n}(x)$ with additional phases.
According to $I_{-n}(z)=I_{n}(z)$, the module of the eigenstate is symmetric about the center.
We compare the numerical and analytical results in Fig.\ref{fig:Q2}.
For $F=0.8$ ($F>F_{c_2}$), the localization length is small and the effect of boundary can be ignored.
Thus both the eigenstates near the boundary and far from the boundary can be described by the analytic results.
The results are shown in Fig.\ref{fig:Q2}(a,b).
For $F=0.1$ ($F<F_{c_2}$), there are some eigenvalues deviating from the imaginary axis (see Fig.\ref{fig:Ksymmetry}(a)), which do not match the analytical results.
The corresponding eigenstates are distributed on the boundary and cannot be described by the analytical results.
A typical eigenstate is shown in Fig.\ref{fig:Q2}(c).
However, eigenstates far from the boundary are not affected by the boundary and can still be well described by the analytical results.
We plot the eigenstate in the middle of the energy spectrum (located in the imaginary axis in Fig.\ref{fig:Ksymmetry}(a)) in Fig.\ref{fig:Q2}(d), which is well described by the analytical result.

\begin{figure}
\centering{}
\includegraphics[scale=0.32]{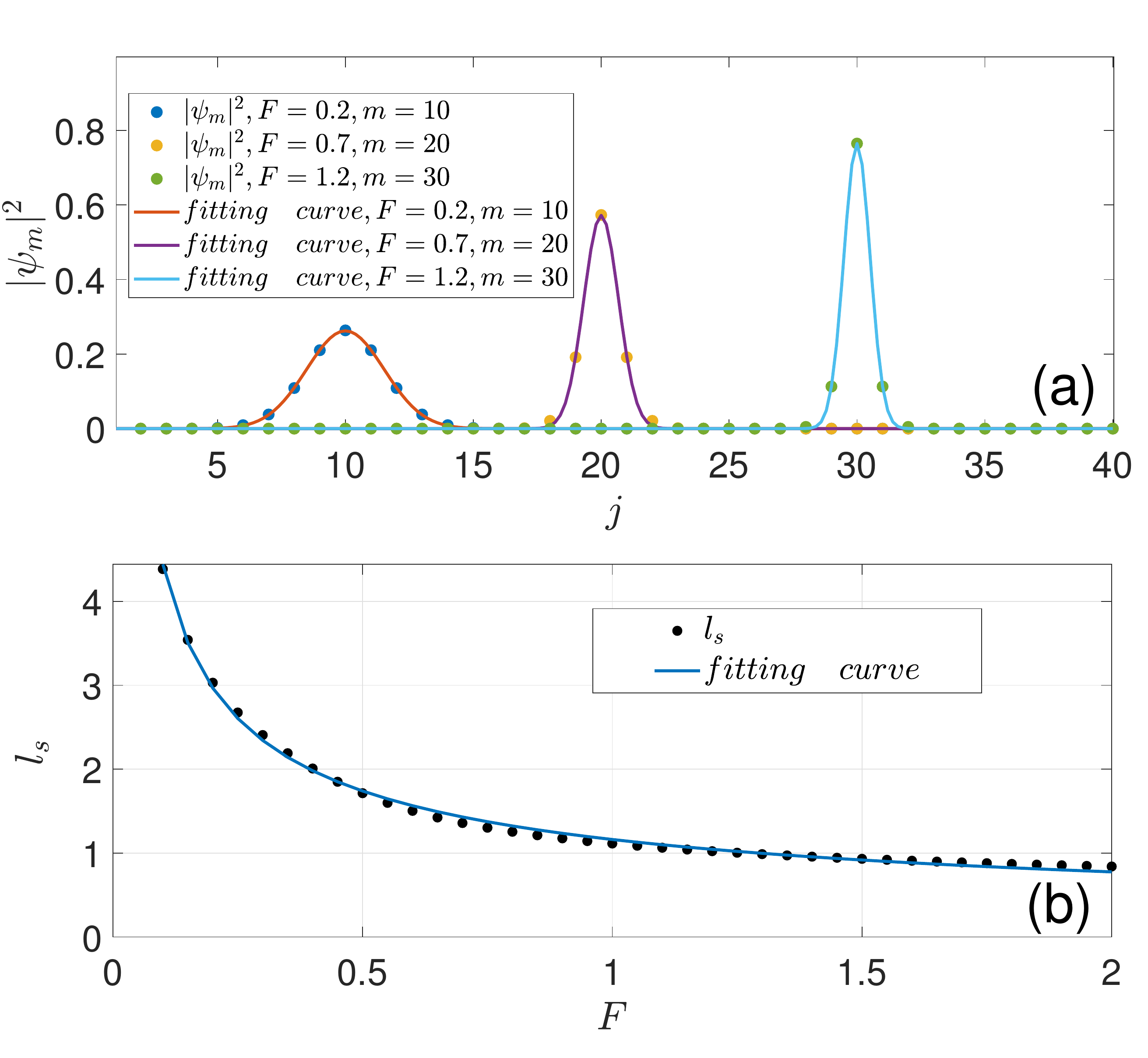}
\caption{(a) The points are numerical results for eigenstates with different $F$ and state label $m$. The curves are fitting results. (b) Localization length $l_s$ versus $F$.
We find the relationship between the $l_s$ and $F$ can be fitted by $l_s \approx 1.162/F^{0.5823} $.  \label{fig:localization_length1}}
\end{figure}

Now we numerically confirm that the eigenstates with different $F$ can be approximated by the Gaussian wavepacket. In Fig.\ref{fig:localization_length1}(a),  we plot $|\psi_m(j)|^2$ versus $j$ with different eigenstate label $m$ for various $F$. The Gaussian wavepackets fit well with the numerical results for all the cases.
The localization length $l_s$ can be also obtained numerically.
In Fig.\ref{fig:localization_length1}(b), we fit the relationship between $F$ and $l_s$ in the region near $F_{c_2}$, where the eigenstates are localized. The numerical results show that they fufill $l_s \propto \frac{1}{F^{\alpha}}$.

\begin{figure}
\centering{}
\includegraphics[scale=0.6]{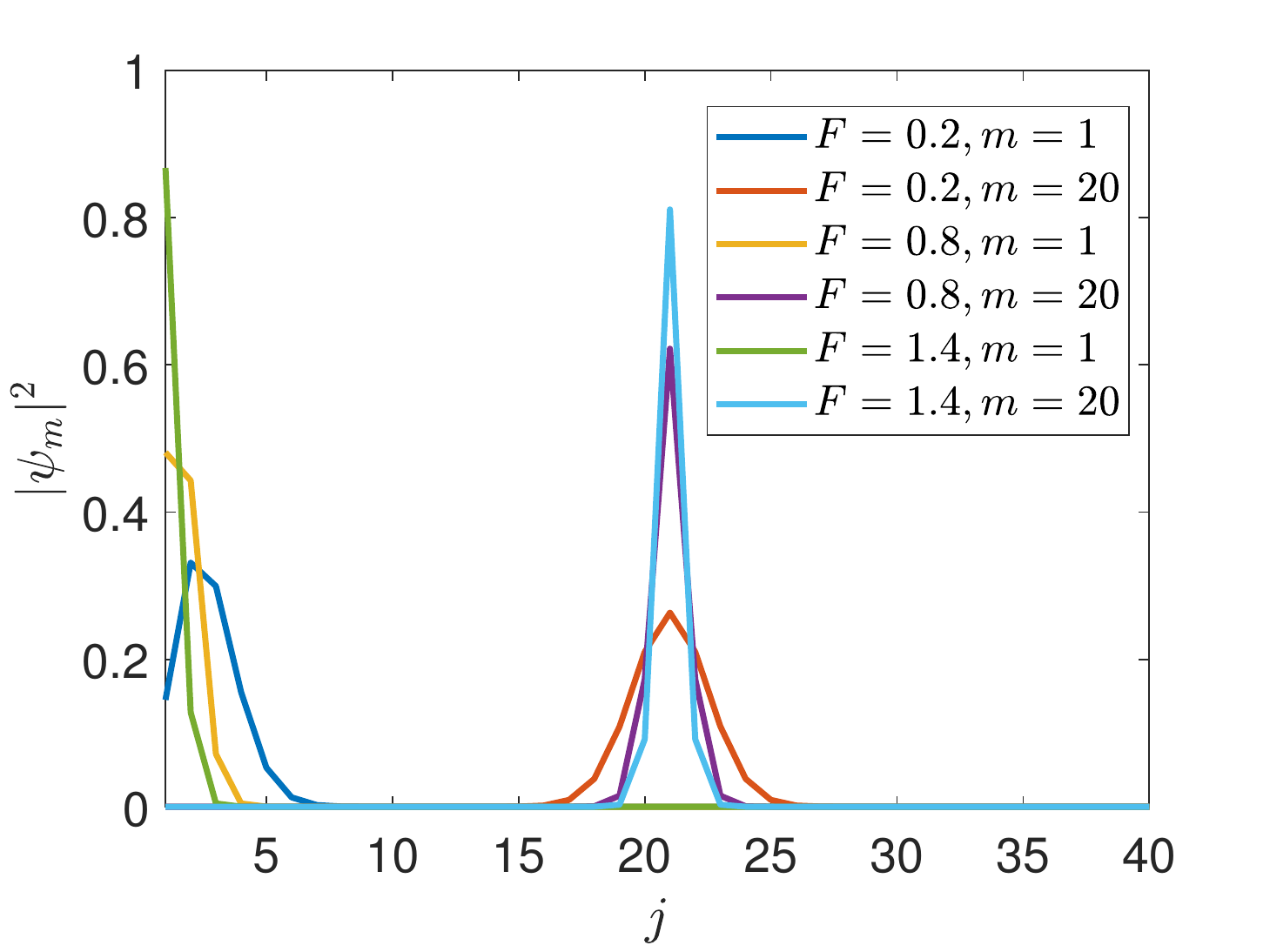}
\caption{$|\psi_m|^2$ for eigenstates near the boundary and far from the boundary with different $F$. }
\label{A3}
\end{figure}

The above discussion mainly focuses on the eigenstates which are far from the boundary and the effect of boundary can be neglected.
Next we study the eigenstates near the boundary and the relationship with $F$.
We sort the eigenstates with the imaginary part of their eigenvalues.
The eigenstates localized near the boundary corresponding to the eigenvalues with the maximum and minimum imaginary parts.
In order to compare the eigenstates near the boundary and far from the boundary, we plot the eigenstates with label $m=1$ and $m=20$ for different $F$ in Fig.\ref{A3}.
All the eigenstates become more localized when we increase $F$.
However, the localization center of the eigenstates move to the boundary gradually for the boundary eigenstates while the eigenstates in the middle region remains at the same position.


\section{Lindblad evolution and non-Hermitian dynamics \label{LindbladVsNonHermitian}}
In this appendix, we discuss the relationship between non-Hermitian dynamics and the Lindblad evolution.
It is known that the dynamics of open quantum system is governed by the effective non-Hermitian Hamiltonian if the jump terms are omitted.
In general, non-Hermitian dynamics is only a short time approximation of Lindblad evolution as the jump terms also contribute to the dynamical behaviors.
Here, we show the equivalence between Lindblad evolution and non-Hermitian dynamics when we calculate the single particle correlation matrix in the single particle case.
For the case that only contains the loss terms, the results obtained by the non-Hermitian dynamics and the Lindblad master equation are the same.
This equivalence can be understood as follows: the particle number of the initial state is a certain number and the initial state density matrix $\rho(0)$ is block diagonalized in particle number space.
The jump terms in the Lindblad master equation ($2L^\dagger_{\mu}\rho L_{\mu}$) have no effects on the dynamics in the subspace of the initial number.
We consider the projection operator $P_N$ which projects the wavefunction to a subspace with $N$ particles and the jump term satisfies: $P_N 2L^\dagger_{\mu}\rho L_{\mu} P_N=0$. That means the dynamics in the subspace with $N$ particles is completely determined by the effective Hamiltonian \cite{NH_LME}.

In our calculation, there are only loss terms and the initial state was chosen in the subspace with $N=1$. The evolution of the density matrix in the subspace $N=1$ can be written as
$$ \rho_{eff}(t)=e^{-i\hat{H}_{eff}t}\rho(0)e^{i\hat{H}_{eff}t}.$$
Then we consider the correlation function $\Delta_{jj'}(t)=Tr[\hat{c}^\dagger_j \hat{c}_{j'} \rho(t)]$ and we only need to consider the subspace $N=1$ and $N=0$.
The single particle correlation function can be written as:
\begin{align*}
& \text{Tr}(\hat{c}^\dagger_j \hat{c}_{j'} \rho(t)) \\
  =& \sum_{j''} \langle j''|\hat{c}^\dagger_j \hat{c}_{j'} \rho(t) |j''\rangle + \langle Vac|\hat{c}^\dagger_j \hat{c}_{j'} \rho(t) |Vac\rangle\\
                           =&\sum_{j''} \langle j''|\hat{c}^\dagger_j \hat{c}_{j'} \rho(t) |j''\rangle =\sum_{j''} \langle j''|\hat{c}^\dagger_j \hat{c}_{j'} \rho_{eff}(t) |j''\rangle,
\end{align*}
where $| Vac \rangle$ represents the vacuum state. So the single particle correlation is determined by the dynamics in the subspace $N=1$ and is equivalent to result in non-Hermitian dynamics.

However, the dynamics given by a non-Hermitian Hamiltonian and the Lindblad master equation have many differences.
One important part is the purity of the density matrix:
the wavefunction in non-Hermitian dynamics can be written as
$$
|\psi(t)\rangle=c(t)\exp(-i\hat{H}_{eff}t)|\psi(0)\rangle,
$$
where $c(t)$ is the normalization coefficient and the wavefunction is always a pure state.
However, the probability to find the vacuum state is always $|\langle Vac|\psi\rangle|^2=0$. It is clear that the non-Hermitian dynamics is not an effective description for the evolution of purity.
As for the Lindblad evolution, the Lindblad equation will lead to the state mixed.
The time evolution of density matrix in our model can be written as
$$
\rho(t)=\rho_{eff}(t)+[ 1-\text{Tr}(\rho_{eff}(t)) ]|Vac \rangle \langle Vac|
$$
and the density matrix in Lindblad evolution contains more information than the non-Hermitian dynamics.

When we consider many-body cases, the Lindblad evolution becomes complicated and the non-Hermitian Hamiltonian can not describe the evolution dynamics effectively.
However, when we focus on the single particle correlation evolution, the evolution is determined by the X matrix \cite{chiraldamping}: $X=iH^* $, which is similar to a single particle Hamiltonian. This holds true even when the initial state is chosen as a many-body state.
In this respect, different properties of X matrix will lead to different behaviors in dynamics.

\section{Some analytic results for the dynamical evolution \label{open dynamical}}
In this part, we give some analytic results for the time evolution in the limit case ($F$ is very small or large).
For a system with dissipative term, the dynamical behavior of the single particle correlation is determined by the equation:
\[
\frac{d\Delta(t)}{dt}=X\Delta(t)+\Delta(t)X^{\dagger},
\]
where $X=ih^{T}-M_{l}=iH^{*}$.
For the dissipative nature, the steady state of our system is an empty state $(\Delta_s(t)=0)$ and the deviation $\tilde{\Delta}(t)=\Delta(t)-\Delta_s=\Delta(t)$.
Then we have
\[
\Delta(t)=e^{Xt}\Delta(0)e^{X^{\dagger}t}.
\]

We can write $X$ matrix in terms of right and
left eigenvectors:
\[
X=\sum_{n}\lambda_{n}|u_{Rn}\rangle\langle u_{Ln}|
\]
and the single-particle correlation can be written as
\[
\Delta(t)=\sum_{m,n}\exp\left[(\lambda_{n}+\lambda_{m}^{*})t\right]|u_{Rn}\rangle\langle u_{Ln}|\Delta(0)|u_{Lm}\rangle\langle u_{Rm}|.
\]
We choose an initial state in which the particles are located at the
center $m_{1}$, i.e., $\Delta_{ij}(0)=\delta_{im_{1}}\delta_{jm_{1}}$.

(i) When $F$ is small (passive-$\mathcal{PT}$-symmetry unbroken), the real
part of the eigenvalues of $X$ matrix are the same. Thus we have
\begin{align*}
\Delta(t) & =\sum_{m,n}\exp\left[(\lambda_{n}+\lambda_{m}^{*})t\right]|u_{Rn}\rangle\langle u_{Ln}|\Delta(0)|u_{Lm}\rangle\langle u_{Rm}|\\
= & \exp\left(Re(\lambda_{n}+\lambda_{m}^{*})t\right) \times \\
 & \sum_{m,n}\exp\left[Im(\lambda_{n}+\lambda_{m}^{*})t\right]|u_{Rn}\rangle\langle u_{Ln}|\Delta(0)|u_{Lm}\rangle\langle u_{Rm}|\\
=& \exp\left(-F(L+1)t  \right) \times \\
 & \sum_{m,n}\exp\left[Im(\lambda_{n}+\lambda_{m}^{*})t\right]|u_{Rn}\rangle\langle u_{Ln}|\Delta(0)|u_{Lm}\rangle\langle u_{Rm}| .
\end{align*}
The dynamical behavior is similar to a Hermitian system but with a global dissipation which is related to  $E_0$.

(ii) When $F$ is large (all the eigenvalues are distributed on the imaginary
axis), all the eigenstates are localized and we can get:
\begin{align*}
\Delta(t) & =\sum_{m,n}\exp\left[(\lambda_{n}+\lambda_{m}^{*})t\right]|u_{Rn}\rangle\langle u_{Ln}|\Delta(0)|u_{Lm}\rangle\langle u_{Rm}|\\
= & \exp\left[2\lambda_{m_{1}}t\right]|u_{Rm_{1}}\rangle\delta_{im_{1}}\delta_{jm_{1}}\langle u_{Rm_{1}}|,
\end{align*}
which means the wavepacket decays rapidly in one lattice
before it propagates to another sites.

Now we explain why the contours of density distribution can be described by hyperbolic curves in the region of $F>F_{c_2}$.
The initial state is chosen as the state with all the sites being occupied. The $\Delta(t)$ can also be written as:
\[
\Delta(t)=\sum_{m,n}\exp\left[\left(\lambda_{n}+\lambda_{m}^{*}\right)t\right]|u_{R,n}\rangle\langle u_{L,n}|\Delta(0)|u_{L,m}\rangle\langle u_{R,m}|.
\]
Then we consider the density distribution as the $j_{th}$ site, which
\begin{align*}
n_j(t)  & = \Delta(t)_{jj}  =\langle j|\Delta(t)|j\rangle\\
 =& \sum_{m,n}\exp\left[\left(\lambda_{n}+\lambda_{m}^{*}\right)t\right]\langle j|u_{R,n}\rangle\langle u_{L,n}|j\rangle\langle u_{R,n}|\Delta(0)|u_{L,m}\rangle.
\end{align*}
We assume $\langle j|u_{R,n}\rangle\approx\delta_{jn}$ and
$\langle u_{R,m}|j\rangle\approx\delta_{mj}$, and it follows
\[
\Delta(t)_{jj}\approx\exp\left[\left(\lambda_{j}+\lambda_{j}^{*}\right)t\right]\langle j|\Delta(0)|j\rangle .
\]
Since the initial state is fully occupied, $\langle j|\Delta(0)|j\rangle$ is the same for all the $j$.
Then we get
\begin{equation}
n_j(t) \approx C_{0}e^{-2Fjt},
\end{equation}
where $C_{0}=\langle j|\Delta(0)|j\rangle=n_j(0)$ is a constant. It follows
\[
\ln n_j(t) \approx \ln C_{0} -2Fjt.
\]
Then a contour with fixed local density $n_c$  fulfills $$t \approx-\frac{\ln(n_c)-\ln C_{0}}{2Fj}.$$


\begin{thebibliography}{10}

\bibitem{Cirac2009} F. Verstraete, M. M. Wolf, and J. Cirac, Quantum computation and quantum-state engineering driven by dissipation, Nat. Phys. \textbf{5}, 633 (2009).

\bibitem{Muller} M. M{\"u}ller, S. Diehl, G. Pupillo, and P. Zoller, Engineered open systems and quantum simulations with atoms and ions, Adv. At. Mol. Opt. Phys. {\bf 61}, 1 (2012).

\bibitem{Barreiro} J. T. Barreiro, M. M{\"u}ller, P. Schindler, D. Nigg, T. Monz, M. Chwalla, M. Hennrich, C. F. Roos, P. Zoller, and R. Blatt, An open-system quantum simulator with trapped ions, Nature {\bf 470}, 486 (2011).

\bibitem{Syassen} N. Syassen, D. M. Bauer, M. Lettner, T. Volz, D. Dietze, J. J. Garc{\'i}a-Ripoll, J. I. Cirac, G. Rempe, and S. D{\"u}rr, Strong dissipation inhibits losses and induces correlations in cold molecular gases, Science {\bf 320}, 1329 (2008).

\bibitem{Diehl2011} S. Diehl, E. Rico, M. A. Baranov, and P. Zoller, Topology by dissipation in atomic quantum wires, Nat. Phys. {\bf 7}, 971 (2011).

\bibitem{LuoL} J. Li, A. K. Harter, J. Liu, L. de Melo, Y. N. Joglekar, and L. Luo, Observation of parity-time symmetry breaking transitions in a dissipative Floquet system of ultracold atoms, Nat. Commun. {\bf 10}, 855 (2019).

\bibitem{Naghiloo} M. Naghiloo, M. Abbasi, Y. N. Joglekar, and K. W. Murch, Quantum state tomography across the exceptional point in a single dissipative qubit, Nat. Phys. {\bf 15}, 1232 (2019).

\bibitem{ZhangW} L. Ding, K. Shi, Q. Zhang, D. Shen, X. Zhang, and W. Zhang, Experimental determination of PT-symmetric exceptional points in a single trapped ion, Phys. Rev. Lett. {\bf 126}, 083604 (2021)

\bibitem{XueP} L. Xiao, K. Wang, X. Zhan, Z. Bian, K. Kawabata, M. Ueda, W. Yi, and P. Xue, Observation of critical phenomena in parity-time-symmetric quantum dynamics, Phys. Rev. Lett. {\bf 123}, 230401 (2019).

\bibitem{TELee}T. E. Lee, Anomalous edge state in a non-hermitian lattice, Phys. Rev. Lett. {\bf 116}, 133903 (2016).

\bibitem{SYao1} S. Yao and Z. Wang, Edge states and topological invariants of non-hermitian systems, Phys. Rev. Lett. {\bf 121}, 086803 (2018).

\bibitem{Kunst} F. K. Kunst, E. Edvardsson, J. C. Budich, and E. J. Bergholtz, Biorthogonal bulk-boundary correspondence in non-hermitian systems, Phys. Rev. Lett. \textbf{121}, 026808 (2018).

\bibitem{Alvarez} V. M. Martinez Alvarez, J. E. Barrios Vargas, and L. E. F.Foa Torres, Non-hermitian robust edge states in one dimension: anomalous localization and eigenspace condensation at exceptional points, Phys. Rev. B {\bf 97}, 121401(R) (2018).

\bibitem{LeeCH} C. H. Lee and R. Thomale, Anatomy of skin modes and topology in non-hermitian systems, Phys. Rev. B {\bf 99}, 201103 (2019).

\bibitem{BL} D. Bernard and A. LeClair, A classification of non-Hermitian random matrices, arXiv:0110649.

\bibitem{Gong} Z. Gong, Y. Ashida, K. Kawabata, K. Takasan, S. Higashikawa, and M. Ueda, Topological phases of non-hermitian systems, Phys. Rev. X \textbf{8}, 031079 (2018).

\bibitem{Sato} K. Kawabata, K. Shiozaki, M. Ueda, and M. Sato, Symmetry and topology in non-Hermitian physics, Phys. Rev. X \textbf{9}, 041015 (2019).

\bibitem{HYZhou} H. Zhou and J. Y. Lee, Periodic table for topological bands with non-Hermitian symmetries, Phys. Rev. B \textbf{99}, 235112 (2019).


\bibitem{LiuCH-2019} C.-H. Liu and S. Chen, Topological classification of defects in non-Hermitian systems, Phys. Rev. B {\bf 100}, 144106 (2019).

\bibitem{LiuCH-PRB} C.-H. Liu, H. Hu and S. Chen, Symmetry and topological classification of Floquet non-Hermitian systems, Phys. Rev. B {\bf 105}, 214305 (2022).

\bibitem{skin effect dissipative}S. Longhi, Unraveling the non-hermitian skin effect in dissipative systems, Phys. Rev. B 102, 201103(R) (2020).

\bibitem{information constraint}C.-H. Liu and S. Chen, Information constraint in open quantum systems, Phys. Rev. B {\bf 104}, 174305 (2021).


\bibitem{chiraldamping}F. Song, S. Yao, and Z. Wang, Non-hermitian skin effect and chiral damping in open quantum systems, Phys. Rev. Lett. {\bf 123}, 170401 (2019).

\bibitem{helicaldamping}C.-H. Liu, K. Zhang, Z. Yang, and S. Chen, Helical damping and dynamical critical skin effect in open quantum systems, Phys. Rev. Research, {\bf 2}, 043167 (2020).

\bibitem{WangZ2022} W.-T. Xue, Y.-M. Hu, F. Song, and Z. Wang, Non-hermitian edge burst, Phys. Rev. Lett. {\bf 128}, 120401 (2022).

\bibitem{YanB} Q. Liang, D. Xie, Z. Dong, H. Li, H. Li, B. Gadway, W. Yi, and B. Yan, Dynamic signatures of non-hermitian skin effect and topology in ultracold atoms, Phys. Rev. Lett. {\bf 129}, 070401 (2022).

\bibitem{LiuYG} P. He, Y.-G. Liu, J.-T. Wang, and S.-L. Zhu, Damping transition in an open generalized Aubry-Andr\'{e}-Harper model, Phys. Rev. A \textbf{105}, 023311 (2022).

\bibitem{YiW} T. Li, Y.-S. Zhang, and W. Yi, Engineering Dissipative Quasicrystals, Phys. Rev. B \textbf{105}, 125111 (2022).

\bibitem{Wannier} G. H. Wannier, Wave functions and effective hamiltonian for bloch electrons in an electric field, Phys. Rev. \textbf{117}, 432 (1960).

\bibitem{stark_localization_wainner} G. H. Wannier, Dynamics of band electrons in electric and magnetic fields, Rev. Mod. Phys. \textbf{34}, 645 (1962).

\bibitem{Dahan} M. B. Dahan, E. Peik, J. Reichel, Y. Castin, and C. Salomon, Bloch oscillations of atoms in an optical potential, Phys. Rev. Lett. {\textbf 76}, 4508 (1996).

\bibitem{Raizen} S. R. Wilkinson, C. F. Bharucha, K. W. Madison, Q. Niu, and M. G. Raizen, Observation of atomic wannier-stark ladders in an accelerating optical potential, Phys. Rev. Lett. {\textbf 76}, 4512 (2016).


\bibitem{stark_localization_review} M. Gl{\"u}cka, A. R. Kolovsky, H. J. Korsch, Wannier--stark resonances in optical and semiconductor superlattices, Phys. Rep. \textbf{366}, 103--182 (2002).

\bibitem{stark-localization-solve} T. Hartmann, F. Keck, H. J. Korsch, and S. Mossmann, Dynamics of bloch oscillations, New J. Phys. \textbf{6}, 2 (2004).



\bibitem{Bloch oscillation}F. Bloch, {\"U}ber die Quantenmechanik der Electronen in Kristallgittern, Z. Phys. \textbf{52}: 555-600 (1928).

\bibitem{Zener tunneling}C. Zener, A theory of electrical breakdown of solid dielectrics, R. Soc. Lond. A \textbf{145}: 523-529 (1934).

\bibitem{stark ac+dc} M. Holthaus and D. W. Hone, Localization effects in ac-driven tight-binding lattices, Philos. Mag. B, \textbf{74:2}, 105-137 (1996).


\bibitem{stark-MBL-theory}M. Schulz, C. A. Hooley, R. Moessner, and F. Pollmann, Stark many-body localization, Phys. Rev. Lett. {\textbf 122}, 040606 (2019).

\bibitem{stark-MBL-experiment}W. Morong, F. Liu, P. Becker, K. S. Collins, L. Feng, A. Kyprianidis, G. Pagano, T. You, A. V. Gorshkov and C. Monroe, Observation of stark many-body localization without disorder, Nature, {\textbf 599}, 393 (2021).


\bibitem{Bloch-oscillation PT symmetry}S. Longhi, Bloch oscillations in complex crystals with $\mathcal{PT}$ symmetry, Phys. Rev. Lett. \textbf{103}, 123601 (2009).

\bibitem{Chiral Zener}S. Longhi, Non-bloch-band collapse and chiral zener tunneling, Phys. Rev. Lett. \textbf{124}, 066602 (2020).

\bibitem{dynamic localization complex crystals}S. Longhi, Dynamic localization and transport in complex crystals, Phys. Rev. B \textbf{80}, 235102 (2009).

\bibitem{non-hermitian skin effect+stark}Y. Peng, J. Jie, D. Yu and Y. Wang, Manipulating non-Hermitian skin effect via electric fields, Phys. Rev. B \textbf{106}, L161402 (2022). 

 \bibitem{PT}   C. M. Bender and S. Boettcher, Real spectra in non-Hermitian Hamiltonians having PT symmetry, Phys. Rev. Lett. \textbf{80}, 5243
(1998).


\bibitem{passivePT} M. Ornigotti and A. Szameit, Quasi $\mathcal{PT}$-symmetry in passive photonic lattices, J. Opt. \textbf{16} 065501 (2014).

\bibitem{passivePT1} Y. N. Joglekar and A. K. Harter, Passive parity-time-symmetry-breaking transitions without exceptional points in dissipative photonic systems, Photon. Res. \textbf{6}, A51-A57 (2018).

\bibitem{ex-lo}Y. X. Liu, Q. Zhou, and S. Chen. Localization transition, spectrum structure, and winding numbers for one-dimensional non-Hermitian quasicrystals, Phys. Rev. B {\bf 104},024201 (2021).

\bibitem{handbook of math}Abramowitz M and Stegun I A 1972 Handbook of Mathematical Functions (New York: Dover).

\bibitem{NH_LME}L. Pan, X. Wang, X. Cui, and S. Chen, Interaction-induced dynamical PT -symmetry breaking in dissipative Fermi-Hubbard models, Phys. Rev. A {\bf 102}, 023306 (2020).

\end{thebibliography}
\end{document}